\documentclass[journal]{new-aiaa}

\usepackage[utf8]{inputenc}
\usepackage{textcomp}
\usepackage{multirow}
\usepackage{doi}

\usepackage{graphicx}
\usepackage{amsmath}
\usepackage[version=4]{mhchem}
\usepackage{siunitx}
\usepackage{booktabs,longtable,multicol,tabularx}
\hypersetup{hidelinks, pdfstartview=FitH}

\usepackage{color}

\title{On Scaling of Hall-Effect Thrusters Using Neural Nets}

\author{Yegor V. Plyashkov\footnote{Engineer; also Research Assistant, Laboratory of Methods for Big Data Analysis, HSE University, 109028, Moscow, Russia; also Master Student, Aerospace Technologies School, Moscow Institute of Physics and Technology, 141701, Moscow, Russia} and Andrey A. Shagayda,\footnote{Senior Researcher}}
\affil{Keldysh Research Centre, 125438, Moscow, Russia}
\author{Dmitrii A. Kravchenko,\footnote{Senior Researcher} and Alexander S. Lovtsov,\footnote{Head of Department}}
\affil{Keldysh Research Centre, 125438, Moscow, Russia}
\author{Fedor D. Ratnikov\footnote{Leading Scientist}}
\affil{HSE University, 109028, Moscow, Russia}

\begin{document}

\maketitle

\begin{abstract}
Hall-effect thrusters (HETs) are widely used for modern near-Earth spacecraft propulsion and are vital for future deep-space missions. Methods of modeling HETs are developing rapidly. However, such methods are not yet precise enough and cannot reliably predict the parameters of a newly designed thruster, mostly due to the enormous computational cost of a HET plasma simulation. Another approach is to use scaling techniques based on available experimental data. This paper proposes an approach for scaling HETs using neural networks and other modern machine learning methods. The new scaling model was built with information from an extensive database of HET parameters collected from published papers. Predictions of the new scaling model are valid for the operating parameters domain covered by the database. During the design, this model can help HET developers estimate the performance of a newly designed thruster. At the stage of experimental research, the model can be used to compare the achieved characteristics of the studied thruster with the level obtained by other developers. A comparison with the state-of-the-art HET scaling model is also presented.
\end{abstract}

\section*{Nomenclature}

{\renewcommand\arraystretch{1.0}
\noindent\begin{longtable*}{@{}l @{\quad=\quad} l@{}}
$D_\text{av}$ & Mean diameter of discharge channel, mm \\
$h$ & Height of discharge channel, mm \\
Type & Thruster type \\
$M$ & Ion mass, Da \\
$\Phi_i$ & Propellant first ionization potential (first ionization cost), eV \\
$U_d$ & Discharge voltage, V \\
$P_d$ & Discharge power, W \\
$I_d$ & Discharge current, A \\
$T$ & Thrust, mN \\
$I_\text{spa}$ & Anode specific impulse, s \\
$I_\text{sp}$ & Total specific impulse, s \\
$\eta_a$ & Anode efficiency \\
$\eta$ & Total efficiency \\
$\dot{m}_a$ & Anode mass flow rate, mg/s \\
$\dot{m}$ & Total mass flow rate, mg/s \\

$g$ & freefall acceleration, m/s$^2$ \\

\multicolumn{2}{@{}l}{Acronyms}\\
APE & Absolute percentage error \\
DC & Direct current \\
FNN	& Feed-forward neural net \\
HET & Hall-effect thruster\\
MAE & Mean absolute error \\
MSE & Mean squared error \\
MS & Hall-effect thruster with magnetic shielding \\
SEM & Semi-empirical model \\
TAL & Hall-effect thruster with anode layer \\
\end{longtable*}}

\section{Introduction}\label{sec:i}

\lettrine{A}{} Hall-effect thruster (HET) is a plasma accelerator in which the gaseous propellant is ionized in a direct current (DC) discharge by electrons drifting in crossed electric and magnetic (ExB) fields. The resulting ions are accelerated in a quasi-stationary electric field \cite{Morozov2000, Kim1998-ro}. Hall thrusters have been developed and used to correct satellite orbits for about half a century \cite{Lev2019-li}. However, no reliable methods have been developed to predict their characteristics \cite{Boeuf2017-vr}. 

The complex physics of the processes in the non-equilibrium magnetized plasma resulting from a gas discharge in crossed ExB fields is the primary reason for the difficulties in reliable HET modeling. Simulating this physics with the accuracy necessary for engineering applications requires a non-stationary three-dimensional fully kinetic numerical model, which would consume enormous computational resources \cite{Kaganovich2020}. Therefore, when designing a new HET, it is common to use a scaling approach, i.e., to predict the output parameters of a new HET based on the experimentally obtained output parameters of some prototype thruster \cite{Morozov1974-vt, A_I_Bugrova_N_A_Maslennikov_A_I_Morozov1991-vr, fruchtman_scaling_laws, E_Ahedo_undated-xh, M_Andrenucci_L_Biagioni_S_Marcuccio_F_Paganucci_and_M_Tobak_undated-xl, J_Ashkenazy_S_Shitrit_and_G_Appelbaum_undated-hh, M_Andrenucci_F_Battista_and_P_Piliero_undated-oh, Daren2005-mi, F_Battista_E_A_D_Marco_and_T_Misuri_undated-xm, Misuri2008-cd, AA_Shagayda_OA_Gorshkov2013-gc, Shagayda2015,Choe2019, Olano_Garcia2020-ds}. 

\begin{figure}[ht!]
\centering
\includegraphics[width=.5\textwidth]{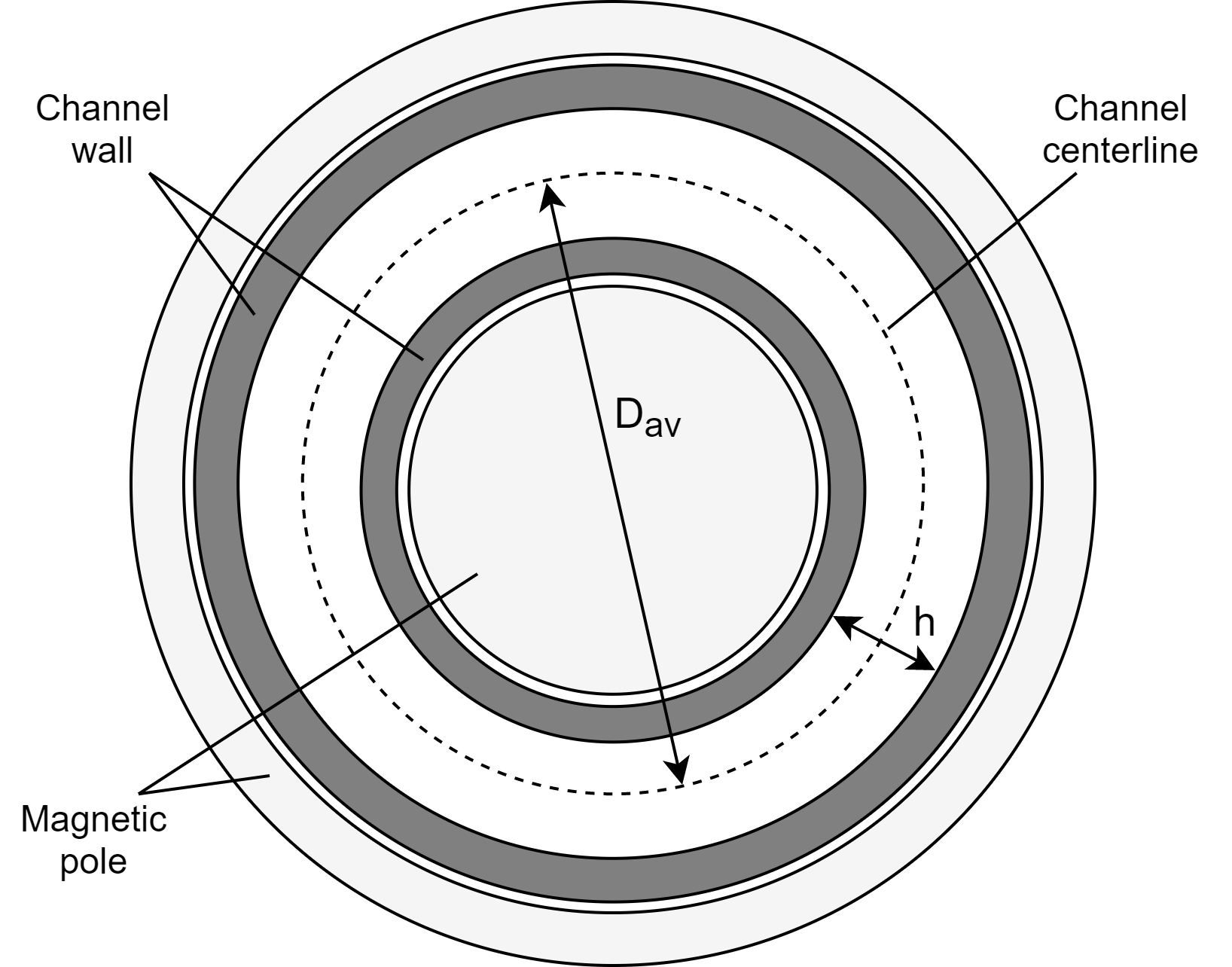}
\caption{Discharge channel mean diameter $D_{av}$ and height $h$.}
\label{fig:HET_scheme}
\end{figure}

The mode of operation of a thruster depends on the following features: discharge voltage $U_d$ and power $P_d$, propellant type and its flow rate $\dot{m}$, discharge channel mean diameter $D_{av}$ and height $h$ (Fig. \ref{fig:HET_scheme}), current in the electromagnets, and others. The thrust $T$ is measured experimentally, which makes it possible to evaluate the specific impulse $I_\text{sp}$ and efficiency $\eta$:
\begin{equation}
\label{Isp and eta}
I_\text{sp} = \frac{T}{\dot{m}}, \quad \eta = \frac{T I_\text{sp}}{2P_d}
\end{equation}

In \cite{Shagayda2015}, a HET scaling approach was developed based on an analysis of a database with details of about 1500 operating modes of more than 30 different thrusters. This model uses only the so-called anode parameters:
\begin{equation}
\label{anode and total parameters}
I_\text{spa} = \frac{\dot{m}}{\dot{m}_a}I_\text{sp}, \quad \eta_{a} = \frac{\dot{m}}{\dot{m}_a} \eta
\end{equation}

It disregards any effects associated with the cathode-neutralizer efficiency and energy losses in the magnetic system. The anode efficiency of a thruster $\eta_a$ can be represented as a product of coefficients responsible for thrust loss in various ways:
\begin{equation}
\label{eta as a product}
\eta_a = \eta_q \eta_m \eta_I \eta_U
\end{equation}
where $\eta_q$ is the charge utilization factor, which penalizes multiply charged ions in the plasma jet,  $\eta_m$ is the propellant utilization factor; i.e., it considers a neutral gas flow, $\eta_I$ is the backstreaming efficiency, which reflects electron leakage, and $\eta_U$ is the non-uniformity or energy efficiency. 

The main idea of the approach is to find semi-empirical physical relations for the discharge parameters and choose values of the parameters in those relations that give the best fit to the available experimental data. As a result, the functional dependencies of thrust, efficiency, and specific impulse on the discharge voltage, power, and the propellant type can be obtained and used to predict the output parameters of a newly developed thruster. 

The model thus constructed has some limitations. First, the only complete model is obtained for the propellant utilization coefficient $\eta_m$. The charge utilization factor $\eta_q$ is approximated using the small amount of experimental data available, which have a large dispersion. The coefficients $\eta_I$ and $\eta_U$ are considered to be fixed.  Second, the gas efficiency model neglects the actual dimensions of the discharge channel. Instead, an unambiguous dependence of the discharge channel size on the discharge power is assumed. In particular, it is assumed that the discharge power determines the optimal mean channel diameter and height. The output characteristics are calculated for these optimal sizes. Therefore, the model developed predicts the operation of a HET with the given discharge channel dimensions but only for a corresponding power value and not for an arbitrarily chosen discharge power range.

In this paper, a new approach for predicting the output characteristics of a HET is proposed based on machine learning algorithms. A more comprehensive database (Sec.~\ref{sec:ii}) is used to train the machine learning models. This database version supplements the previous one with new features (mean diameter and height of the discharge channel, and thruster’s type) on the one hand and new thrusters on the other. There are several advantages of this approach. First, it removes the restrictions associated with poorly substantiated assumptions about the dependency of individual efficiency parameters on the HET operating mode. Second, it simplifies the modeling procedure, thanks to an advanced set of libraries including out-of-the-box machine learning algorithms and tools for preprocessing and analyzing the data, for training, and for estimating the quality of the trained models.

The rest of the paper is structured as follows. Section~\ref{sec:ii} describes the structure of the database, how it was collected, and its statistical characteristics. Section~\ref{sec:iii} describes the machine learning techniques used to build the new scaling model. In Sec.~\ref{sec:iv}, the new model is compared to the state-of-the-art model \cite{Shagayda2015}, and the results obtained are discussed.

\section{Database}\label{sec:ii}

The database used in this study extends the database assembled back in 2012, which was used to build a semi-empirical model (SEM) \cite{Shagayda2015}. The old database consisted of 1527 unique operating modes taken from more than 40 publications covering the period from 1993 to 2012. In total, that database  includes information on about 40 HETs. See the extensive bibliography in \cite{Shagayda2015}.

The database used in this study extends the old one by including information published mainly between 2012 and 2020, such as papers from International Electric Propulsion Conferences and the American Institute of Aeronautics and Astronautics conferences, Ph.D. theses, and other publications. This adds 958 new data points, corresponding to 46 thrusters.

%
                                      %
                                      %
                                      %
%
%
                                      %

\begin{table}[ht!]
\caption{\label{tab:database_structure} Database structure}
\centering
\begin{tabular}{p{5cm} ll}
\toprule
Parameter & Symbol& Units\\
\midrule
\multicolumn{3}{l}{Thruster Construction} \\
   Mean diameter of discharge channel&$D_\text{av}$& mm \\
                                      Height of discharge channel& $h$&  mm \\
                                      Thruster type& Type& -- \\
\\                                                                
\multicolumn{3}{l}{Propellant Parameters} \\
   Ion mass&$M$& Da\\
                  Propellant first ionization potential \newline (first ionization cost) &$\Phi_i$&eV\\                             
\\
\multicolumn{3}{l}{Discharge parameters} \\
   Voltage&$U_{d}$&V\\
                                     Power& $P_d$& W\\                  
\\
\multicolumn{3}{l}{Performance} \\
   Thrust&$T$&N\\
                                    Anode specific impulse&$I_\text{spa}$&s\\
                                    Anode efficiency& $\eta_a$& --\\
\\
\multicolumn{3}{l}{Supplementary information} \\
   Thruster name&--&--\\
                                     Source& --&--\\             
\bottomrule
\end{tabular}
\end{table}

\begin{figure}[ht!]
\centering
\includegraphics[width=1.\textwidth]{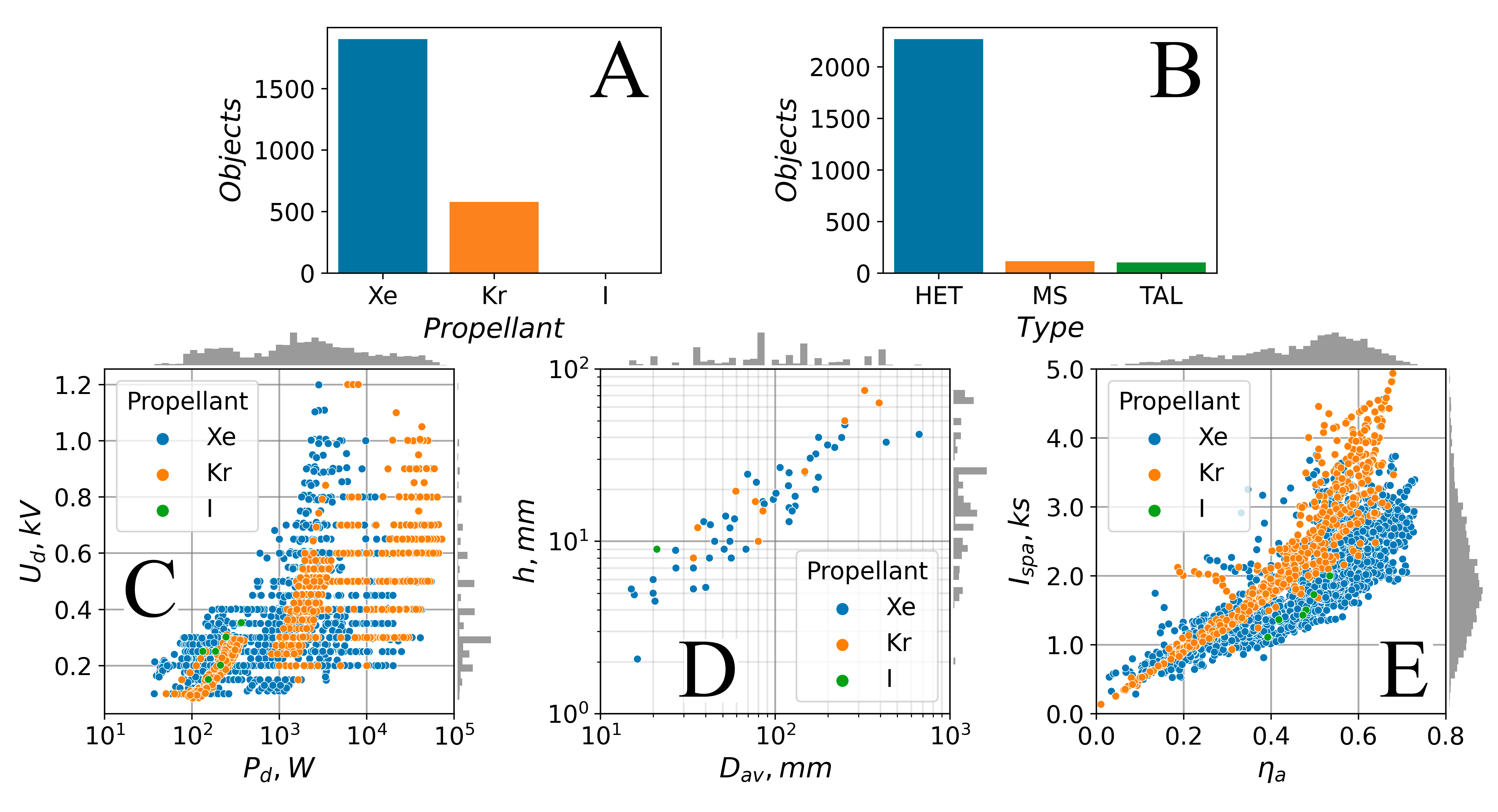}
\caption{Distribution of database entries by propellant type (A), thruster type (B), discharge parameters (C), construction parameters (D), and performance (E).}
\label{fig:dataset_scattering}
\end{figure}

For each discharge mode, the database contains the parameters listed in Table~\ref{tab:database_structure}. Figure~\ref{fig:dataset_scattering} shows the distribution of database entries by propellant type (A), thruster type (B), discharge voltage and power (C), height and mean diameter of the discharge channel (D), and anode specific impulse and efficiency (E).
The database contains information on operating modes for xenon, krypton, and iodine propellants. The number of iodine entries is significantly smaller than for xenon or krypton (Fig.~\ref{fig:dataset_scattering}A).
The construction of a thruster is characterized with the mean diameter and height of the discharge channel as well as the type of thruster. There are three types: (1) HETs, classic Hall-effect thrusters with dielectric discharge channel walls; (2) TALs, thrusters with an anode layer, differing in the conducting walls of the discharge channel; and (3) MSs, thrusters with magnetic shielding, in which the profile of the discharge channel walls and the topology of the magnetic field are chosen to reduce the interaction of the plasma with the surface of the discharge channel walls. The amount of data available for HET operating modes significantly exceeds the amount of data for MS and TAL thrusters (Fig.~\ref{fig:dataset_scattering}B).
Note, that the distribution of the discharge voltage (Fig.~\ref{fig:dataset_scattering}C) as well as those for the mean diameter and height of the discharge channel (Fig.~\ref{fig:dataset_scattering}D) are strongly uneven and discrete. The distributions for the discharge power (Fig.~\ref{fig:dataset_scattering}C), specific impulse, and efficiency (Fig.~\ref{fig:dataset_scattering}E) are smoother. 
The implications of these irregularities in the data distribution are described in Sec.~\ref{sec:iii}. 

Some publications used to compile the database contain only an incomplete set of the parameters required. For example, a significant number of articles provide total operation parameters, which include the effects of a decrease in the efficiency and specific impulse associated with power losses in the cathode and magnetic system.  In some papers the ratio of the total and anode propellant flow rates is specified. We could then calculate the anode output parameters using Eq.~(\ref{anode and total parameters}). If the ratio was not available in a publication, the cathode flow rate was taken to be 10$\%$ of the total, thus making it possible to approximate the anode parameters.

Some publications lacked information on the geometry of the discharge channel. To compensate for this, the channel dimensions were estimated from published images of the thruster. It is also worth mentioning that a significant number of publications contain information about the performance of the truster in graphical rather than tabular form. In this case, some special tools that enable us to extract data points from a graph image were used.

\section{Approach to scaling}\label{sec:iii}

Most models for HET scaling \cite{Morozov1974-vt, A_I_Bugrova_N_A_Maslennikov_A_I_Morozov1991-vr, fruchtman_scaling_laws, E_Ahedo_undated-xh, M_Andrenucci_L_Biagioni_S_Marcuccio_F_Paganucci_and_M_Tobak_undated-xl, J_Ashkenazy_S_Shitrit_and_G_Appelbaum_undated-hh, M_Andrenucci_F_Battista_and_P_Piliero_undated-oh, Daren2005-mi, F_Battista_E_A_D_Marco_and_T_Misuri_undated-xm, Misuri2008-cd, AA_Shagayda_OA_Gorshkov2013-gc, Shagayda2015, Choe2019, Olano_Garcia2020-ds} use various theoretical or semi-empirical approaches. In this work, regular machine learning models were built with the information in the database.
The main difference of the proposed approach is the use of a generic non-parametric model to describe the available data, in contrast to the physics-motivated parametric models used in previous approaches. As a result, the model implicitly learns the empirical patterns in the data, and thus, it can subsequently reproduce these patterns for different input parameter sets.
In this work, a few regressive machine learning models are studied, namely, feed-forward neural networks (FNN) and ensembles of FNN models. Note that supervised machine learning models, such as FNNs \cite{IEPC-2017-453}, recurrent neural networks \cite{IEPC-2019-501, IEPC-2019-169}, and convolution neural networks \cite{IEPC-2019-169}, have already been applied to electric propulsion problems. 

The input parameters of the studied FNN models are the propellant properties ($M$ and $\Phi$), thruster construction ($D_\text{av}$, $h$ and $Type$), and the discharge parameters ($U_d$ and $P_d$). In typical HET tests, the discharge mode is adjusted by voltage and power, defined by the power supply. Power is maintained by tuning the propellant mass flow rate. Therefore, the mass flow rate is usually not an independent feature. Its value, as well as the values of thrust and specific impulse, are determined in tests. We chose the discharge power $P_d$ and voltage $U_d$ as input parameters on this basis.

The anode efficiency $\eta_a$ and anode specific impulse $I_\text{spa}$ were chosen as the output parameters of the models because they vary within only one or two orders of magnitude. Experience has shown the predictions of FNN models are less accurate when the output parameters are distributed over more orders of magnitude. That is why the anode efficiency $\eta_a$ and anode specific impulse $I_\text{spa}$ are more suited to be the output of the studied models than the thrust $T$ and anode propellant mass flow rate $\dot{m}_a$. Furthermore, the thrust $T$ and anode propellant mass flow rate $\dot{m}_a$ can be calculated from $I_\text{spa}$ and $\eta_a$ at a given discharge power $P_d$ according to Eqs.~(\ref{Isp and eta}) and~(\ref{anode and total parameters}).

\begin{figure}[ht!]
\centering
\includegraphics[width=.5\textwidth]{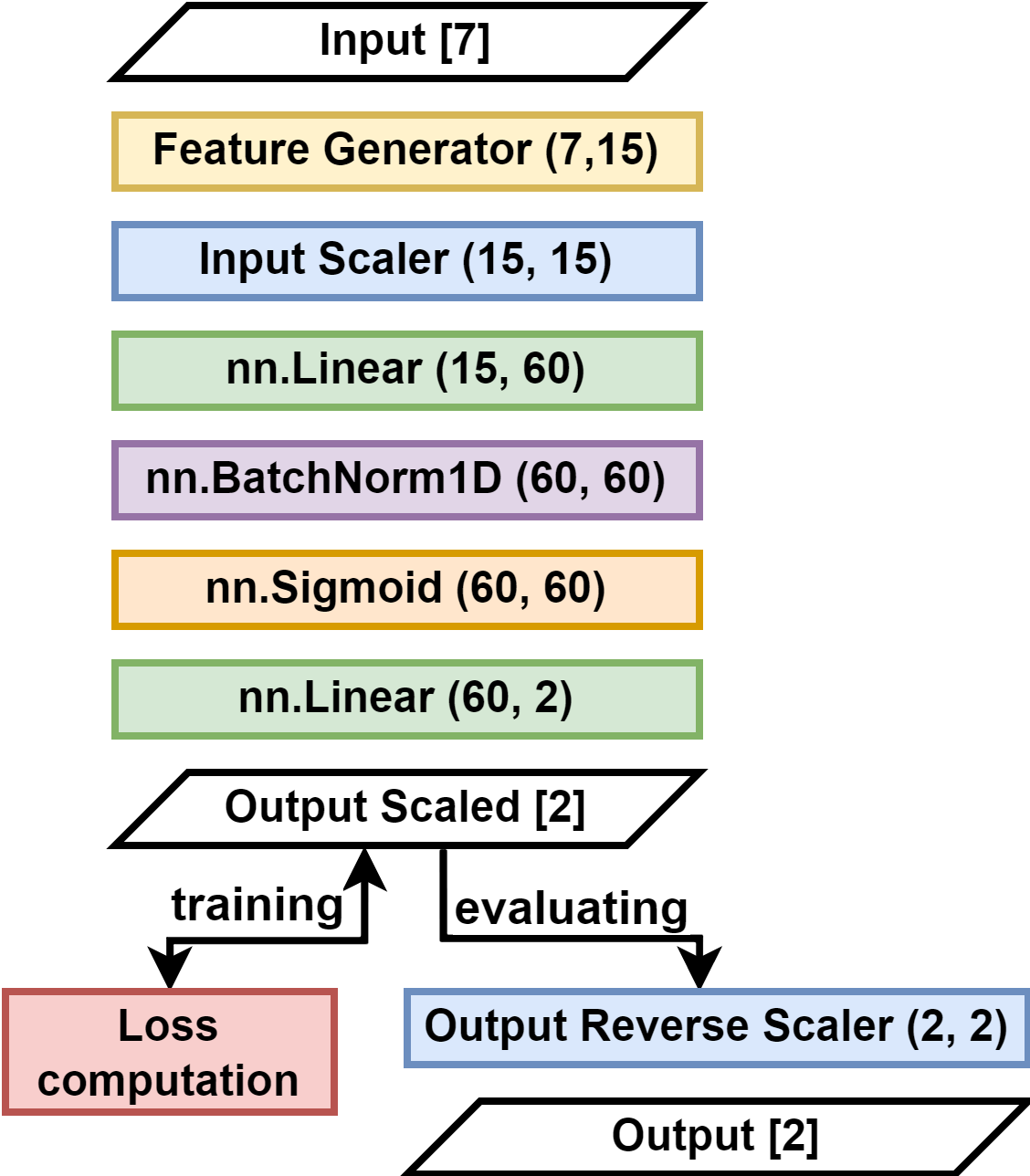}
\caption{Architecture of the feed-forward neural net (FNN).}
\label{fig:FNN}
\end{figure}

The FNN, or multi-layer perceptron, used in this work consists of the sequence of layers presented in Fig.~\ref{fig:FNN}: the additional feature generator, the scaling layer for the input parameters (input scaler), the first fully connected layer (nn.Linear), the batch normalization layer (nn.BatchNorm1D), the nonlinear layer (nn.Sigmoid), the second fully connected layer (nn.Linear), and the scaling layer for the output parameters (output reverse scaler). Layers with the "nn." prescript belong to the PyTorch \cite{Paszke2019} framework for the Python programming language. In Fig.~\ref{fig:FNN}, each layer is represented as a colored rectangle block consisting of the layer name and the brackets containing the dimensions of the input and output parameter set. Colorless rhomboids correspond to inputs or outputs with the dimensions indicated in square brackets.
The only trainable layers in this model are the two linear layers and the batch normalization layer. The other layers are all predefined and not trainable.

\begin{figure}[ht!]
\centering
\includegraphics[width=.5\textwidth]{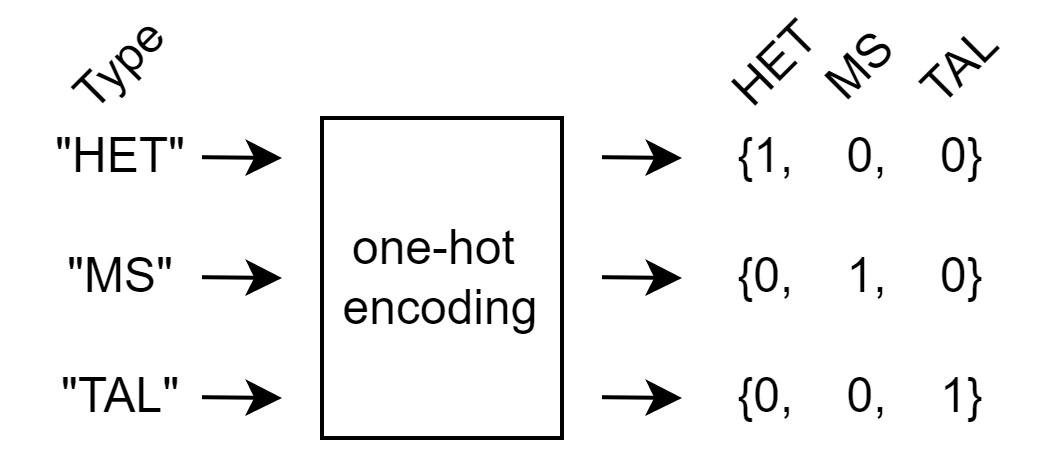}
\caption{One-hot encoding of the $Type$ feature.}
\label{fig:onehot}
\end{figure}

\begin{figure}[ht!]
\centering
\includegraphics[width=0.5\textwidth]{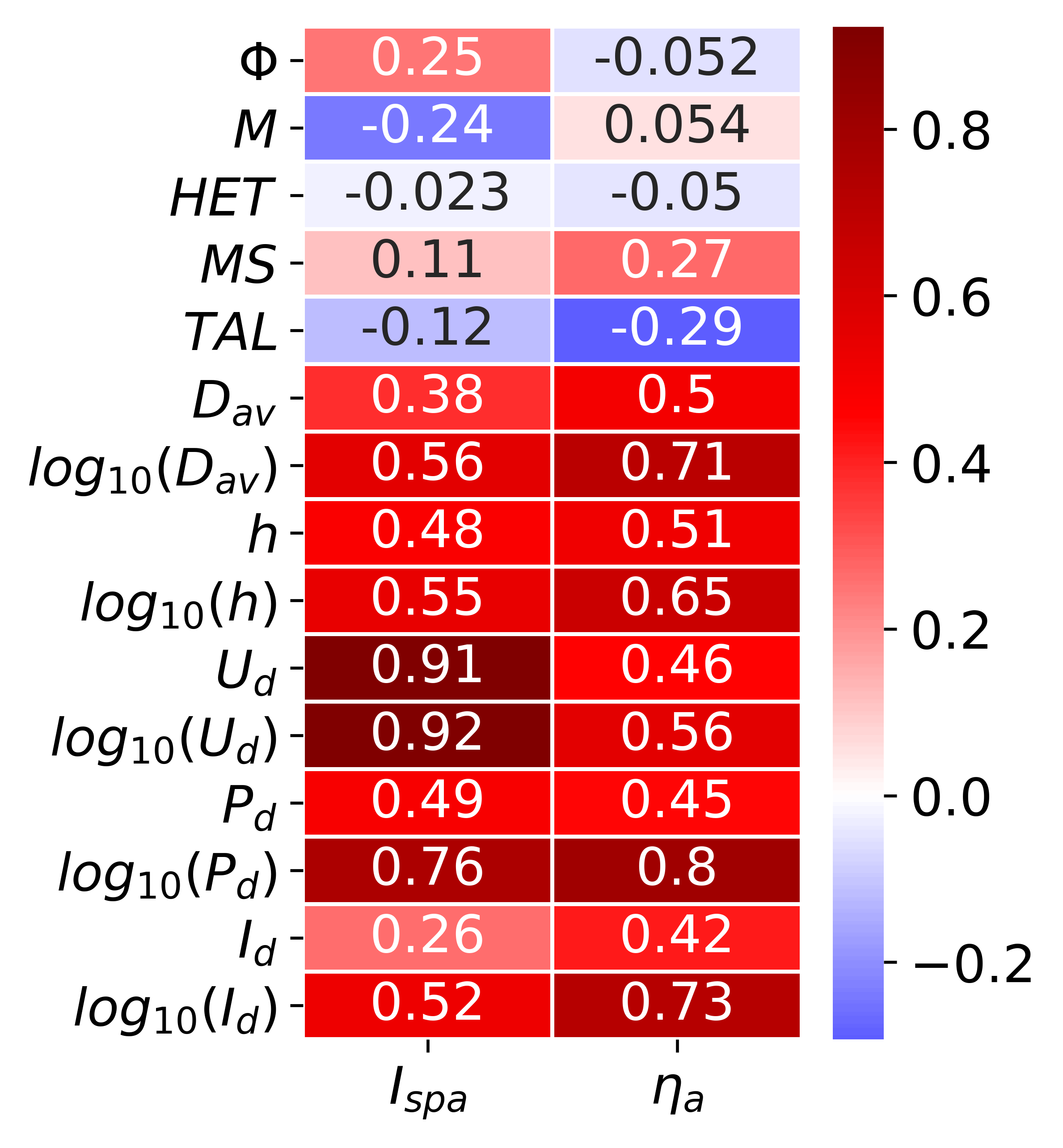}
\caption{Correlation matrix of inputs and outputs.}
\label{fig:corrmat}
\end{figure}

The Feature Generator layer extends the original input features with a few derived ones: the discharge current ($I_d = P_d / U_d$) and the decimal logarithms of $D_\text{av}$, $h$,  $U_d$, $P_d$, and $I_d$. As a result of one-hot encoding, the string representation of the type feature is converted into a three-dimensional vector \{HET, MS, TAL\}, in which the component corresponding to the thruster type is set to 1, and the other two are set to 0 (Fig.~\ref{fig:onehot}). The output of this layer is, thus, a vector of dimension 15, composed of the six original input parameters and 9 derived ones (Table~\ref{tab:FG layer output}). These additional features help to improve the model accuracy because some logarithms of input features correlate with output parameters better than the input features themselves (Fig.~\ref{fig:corrmat}). For instance, $I_{spa}$ and $\eta_a$ correlate better with $log_{10}(P_d)$ than $P_d$. Despite weak correlations with the output parameters, the propellant parameters and one-hot representation of construction type were chosen as input features, since there was an intention to see the difference in predictions for different propellants and thruster types.

\begin{table}[hbt!]
\caption{\label{tab:FG layer output} Output of the Feature Generator layer}
\centering
\begin{tabular}{@{}cccccccccc@{}}
\toprule
Index & 1 & 2 & 3 & 4 & 5 & 6 & 7 & 8 & 9 \\
Feature & $\Phi$ & $M$ & $HET$ & $MS$ & $TAL$ & $D_{av}$ & $log_{10}(D_{av})$ & $h$ & $log_{10}(h)$ \\
 &  &  &  &  &  &  &  &  &  \\
Index & 10 & \multicolumn{2}{c}{11} & 12 & \multicolumn{2}{c}{13} & 14 & \multicolumn{2}{c}{15} \\
Feature & $U_d$ & \multicolumn{2}{c}{$log_{10}(U_d)$} & $P_d$ & \multicolumn{2}{c}{$log_{10}(P_d)$} & $I_d$ & \multicolumn{2}{c}{$log_{10}(I_d)$} \\ \bottomrule
\end{tabular}
\end{table}

The input features are scaled because they naturally have different units and ranges. Scaling allows linear operations on these parameters. The Input Scaling layer performs this according to Eq.~(\ref{input_scaling}):
\begin{equation}
\label{input_scaling}
x_{scaled} = \frac{x - \operatorname{Mean}[X]}{\operatorname{Var}[X]},
\end{equation}
where $x$ refers to the input of this layer, and $x_{scaled}$ to the output, whereas $\operatorname{Mean}[X]$ is the mean and  $\operatorname{Var}[X]$ is the variance of the input feature in the feature array $X$ of the database.

During model training, the loss function is calculated and used to update the weights of the model so that it makes predictions with higher precision. The loss function depends on two output parameters, the anode efficiency $\eta_a$ and anode specific impulse $I_\text{spa}$. Since these parameters originally have different dimensions, they may affect the loss function differently. Thus, the training configuration (Fig.~\ref{fig:FNN}, training double arrow) and evaluation  configuration (Fig.~\ref{fig:FNN}, evaluating arrow) of the FNN are separate with different penultimate layers. In the training configuration, the fully connected layer nn.Linear is the penultimate layer, whereas for evaluation, the output parameters have a scaling layer (output reverse scaler), which operates according to Eq.~(\ref{reverse_scaling}):
\begin{equation}
\label{reverse_scaling}
y = y_{scaled} \operatorname{Var}[Y] + \operatorname{Mean}[Y]
\end{equation}
The input of this layer is the $y_{scaled}$, the output is the target variable $y$ in its original scale and dimension, the $\operatorname{Mean}[Y]$ is the mean and the $\operatorname{Var}[Y]$ is the variance of the target variable $y$ in the target array $Y$ of the database. When training the FNN, the loss function is calculated with dimensionless and normalized representations of the output parameters. When making a prediction, the output parameters are used in their natural scale and dimension.

Since the nonlinear layer of the FNN contains the saturating sigmoid function, it is worth using the batch normalization layer. According to \cite{ioffe2015}, it accelerates the learning process and makes it less sensitive to the initial values of the FNN's learnable parameters. This layer can be described by the Eq.~(\ref{batchnorm}):
\begin{equation}
\label{batchnorm}
y=\frac{x-\operatorname{Mean}[x]}{\sqrt{\operatorname{Var}[x]+\epsilon}} \gamma+\beta,
\end{equation}
where $x$ stands for the input array and the $y$ for the output array. The mean and standard deviation are calculated per-dimension over the mini-batches and $\gamma$ and $\beta$ are learnable parameter vectors of the same size as the $x$ and $y$. The $\epsilon$ is a value added to the denominator for numerical stability. We used its default value of $10^{-5}$.

Many different loss functions are used in modern machine learning. In particular, the mean-squared error (MSE) and mean absolute error (MAE) are the most common in supervised learning. The MAE loss function was used in this work to train the FNN.
Many of the thrusters in the database have parameters that are outliers.
Minimizing the MAE corresponds more to minimizing the median error, whereas minimizing the MSE minimizes the mean error, which is significantly more affected by outliers than the median. Thus, the choice of MAE made it possible to produce a general picture in comparison with MSE by ignoring outliers.

\begin{figure}[ht!]
\centering
\includegraphics[width=.7\textwidth]{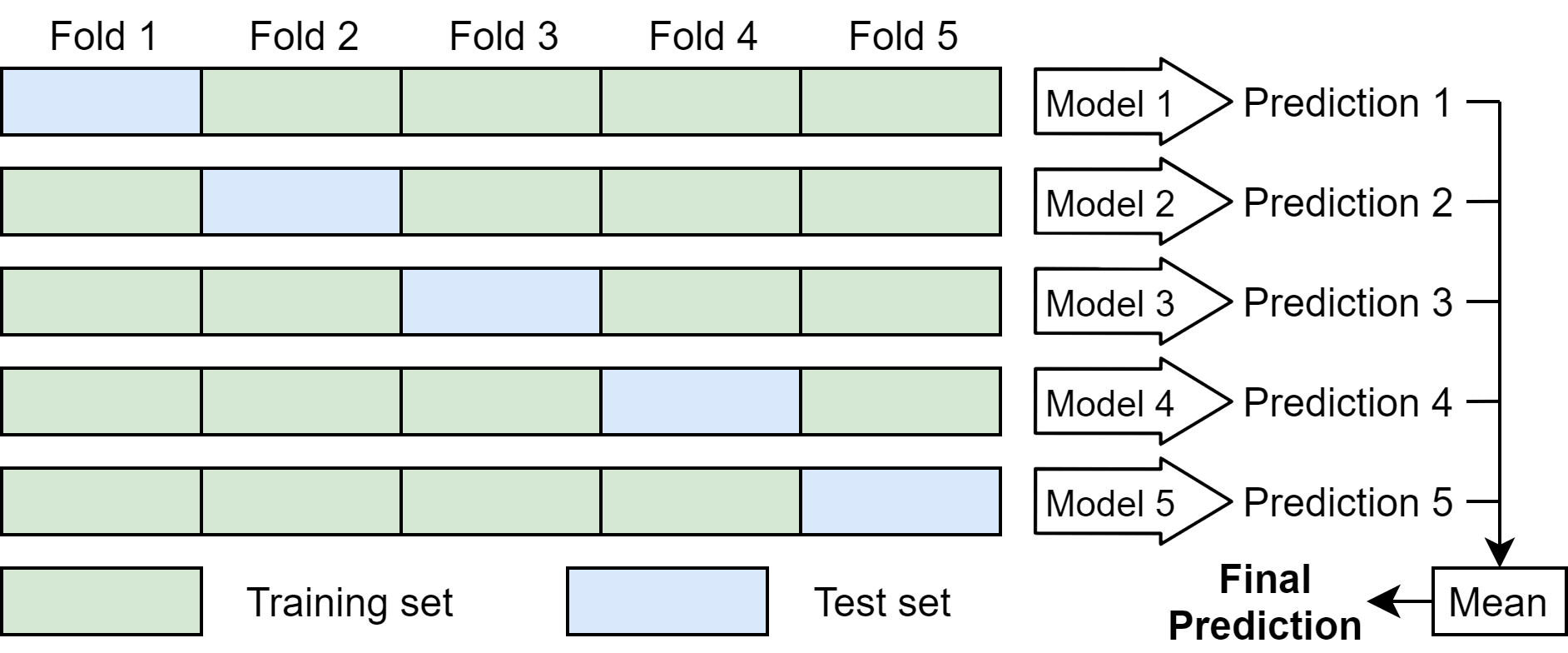}
\caption{Formation of test and training samples with the GroupKFold method.}
\label{fig:ensemble}
\end{figure}

We have noted that the behavior of the trained model depends on the compositions of the training and test samples. To minimize this effect, we applied the GroupKFold method from the Scikit-learn library \cite{Pedregosa2011-nh} and divided the database into five disjoint groups of thrusters (Fig~\ref{fig:ensemble}). The thrusters that filled each group were determined entirely randomly. If a thruster is included in one group, the corresponding discharge modes might be included only in this set but not in the other one. Five pairs of training and test samples were created from these groups, used to train and test five separate models. The average prediction of the five models (or ensemble of those five models) was used for the overall prediction for the thruster parameters.

\begin{figure}[ht!]
\centering
\includegraphics[width=0.5\textwidth]{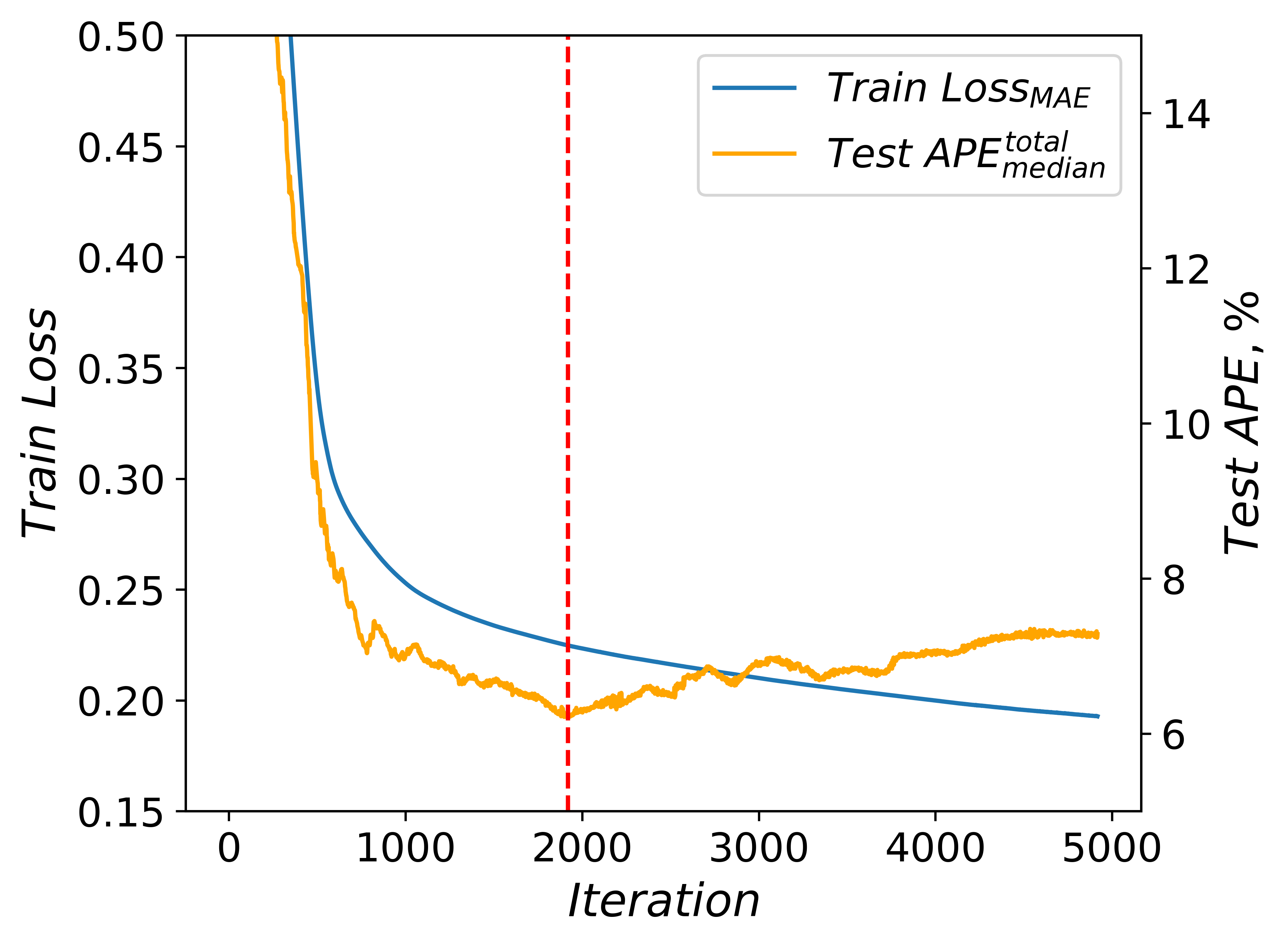}
\caption{Learning curves during the process of training the one of the FNN models.}
\label{fig:learning_curves}
\end{figure}

The Adam stochastic optimizer from the PyTorch framework \cite{Paszke2019} was used to minimize the MAE loss function on the training sample. In the training loop, the model's quality was estimated as a median of the $APE_{total}$ array, which contained both $I_{spa}$ and $\eta_a$ absolute percentage error arrays (Eq.(\ref{APE})) for the test sample. When this value was detected not improving over 3000 iterations, the training process was stopped (Fig. \ref{fig:learning_curves}). After that, the FNN's weights were returned to the state corresponding to the lowest achieved total median APE on the test sample.

\begin{equation}
\label{APE}
\operatorname{APE}_{I_\text{spa}} = \left| \frac{I_\text{spa}^\text{predicted} - I_\text{spa}^\text{true}}{I_\text{spa}^\text{true}} \right| \times 100 \% , \quad
\operatorname{APE}_{\eta_{a}} = \left| \frac{\eta_{a}^\text{predicted} - \eta_{a}^\text{true}}{\eta_{a}^\text{true}} \right| \times 100 \%, \quad \operatorname{APE}_{total} = \{ \operatorname{APE}_{I_\text{spa}},  \operatorname{APE}_{\eta_{a}}\}
\end{equation}

To estimate the performance of the whole ensemble, its accuracy should be calculated for those thrusters that were not used to train it. However, the ensemble, as an average of the five models, is trained on the entire database. Thus, the quality of the ensemble is estimated by averaging the errors of each model for the corresponding test sample.

\begin{figure}[ht!]
\centering
\includegraphics[width=.5\textwidth]{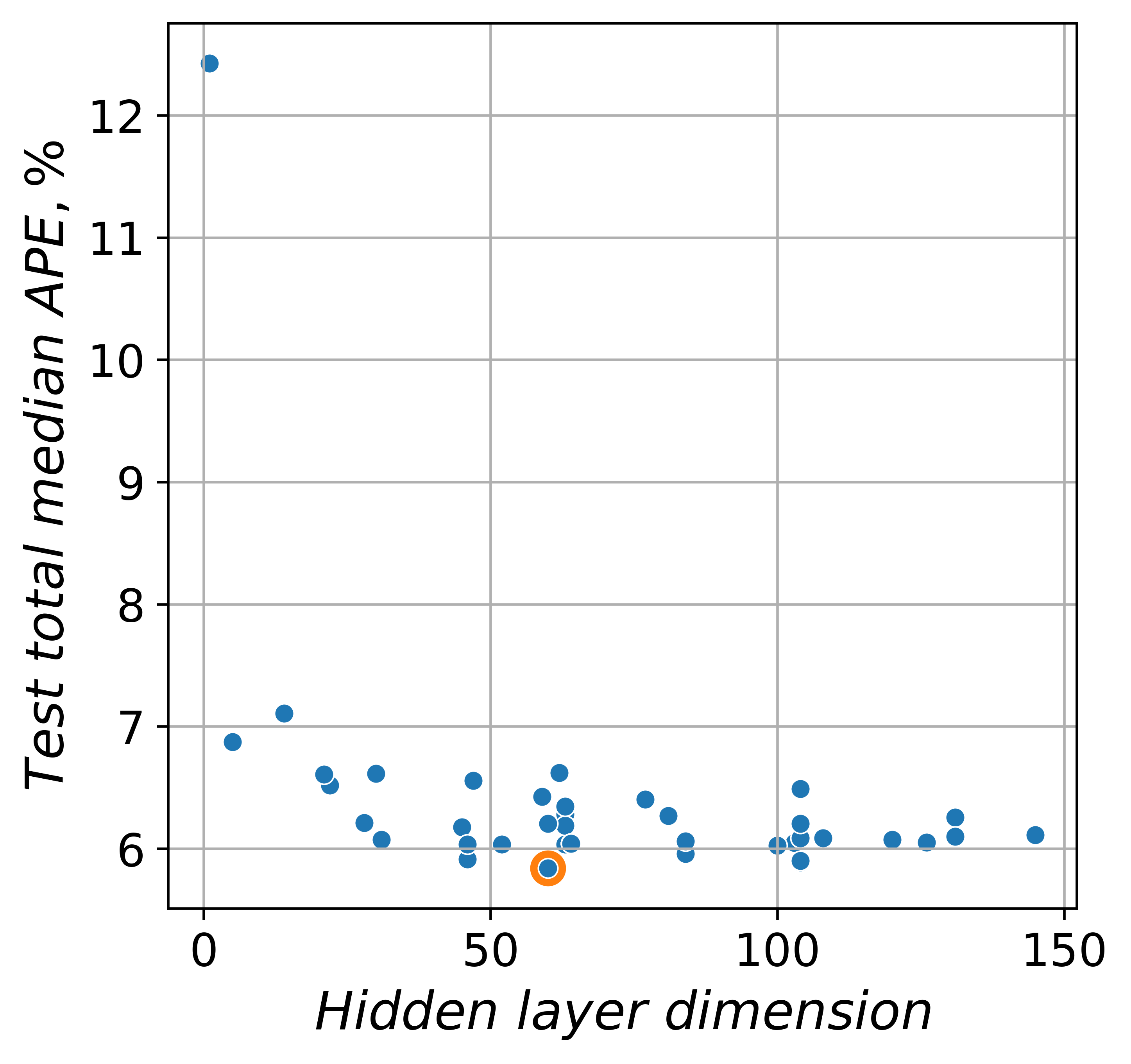}
\caption{Hidden layer dimension optimization.}
\label{fig:optimization}
\end{figure}

One of the parameters affecting the quality of the model is the dimension of its hidden layers, including two linear layers, the Sigmoid, and the BatchNorm layer. The larger the dimension of these layers, the more complex dependencies the model can approximate. However, a model with too many neurons is easier to overfit. At the same time, the model with too few neurons is easy to underfit. The accuracy of the model's predictions for test samples decreases in these cases. Therefore, we trained the FNN ensemble several times for different values of its neural networks' hidden layer dimension. It varied from 1 to 150 during this optimization. Fig.~\ref{fig:optimization} shows that the test median APE is not changing significantly after the value of 45. However, we chose the configuration with the hidden layer dimension of 60 (Fig.~\ref{fig:FNN}) because it provided minimal test median APE.

As mentioned in Sec.2, the number of experimental points with voltage above 650 V in the database is small. Thus, their influence on the model is insufficient. In machine learning, this effect is called data imbalance. To make the data from this domain more influenceable, we increased its size by duplicating randomly-selected domain entries until their number is equal to the required one. We increased the number of modes with discharge voltage between 650V and 950V two times and ones with discharge voltage above 950V twenty times this way. As a result, the tail of the voltage distribution became thicker. The distribution of discharge channel geometry parameters was balanced via the random oversampling method from the imbalanced-learn library \cite{10.5555/3122009.3122026}. This balancing made it possible to obtain more plausible predictions from the trained models for configurations with high voltages or rare constructions.

\section{Results and discussion}\label{sec:iv}

The approach described in Sec.~\ref{sec:iii} was then applied to the data presented in Sec~\ref{sec:ii}. Different aspects of the results obtained and a comparison with the  SEM \cite{Shagayda2015} are discussed in the following sections.

\subsection{Average model precision}

The average model precision is defined as the average relative deviation of the predicted values from those values extracted from the database, based on the APE in Eq.~(\ref{APE}). For the FNN ensemble, this was obtained by averaging those deviations obtained for the five constituent neural networks for the corresponding test samples. For the SEM \cite{Shagayda2015}, the result was obtained by direct averaging of relative deviations for all database entries.

\begin{figure}[htb!]
\centering
\includegraphics[width=1.\textwidth]{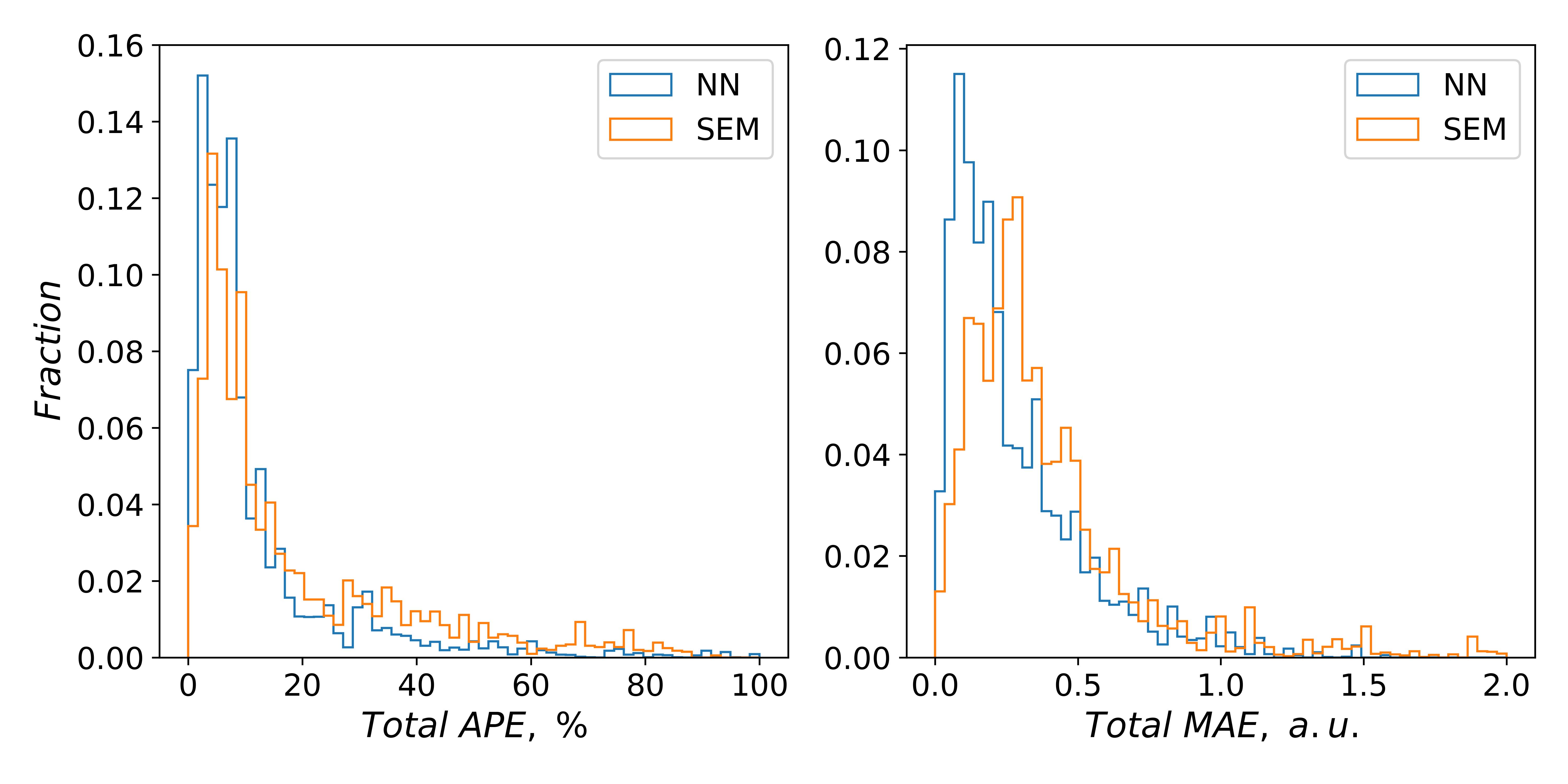}
\caption{Distribution of the total errors for the FNN ensemble and SEM \cite{Shagayda2015}.}
\label{fig:error_distribution}
\end{figure}

Figure ~\ref{fig:error_distribution} shows histograms of the total APE (left) and total MAE (right) distributions for the FNN ensemble (blue) and SEM \cite{Shagayda2015} (orange). To calculate the MAE, the anode efficiency $\eta_a$ and anode specific impulse $I_{spa}$ were scaled in the same way as was done for training the FNN ensemble. Both APE and MAE distributions show that there are more elements with a high error in the predictions of the SEM \cite{Shagayda2015}. Table~\ref{tab:APE_FNN_SEM_comparison} displays the total mean and total median errors corresponding to the total APE distribution for the FNN ensemble and SEM \cite{Shagayda2015}.

\begin{table}[hbt!]
\caption{\label{tab:APE_FNN_SEM_comparison} Absolute percentage errors, $\%$}
\centering
\begin{tabular}{cccclccc}
\toprule
 & \multicolumn{3}{c}{Mean} &  & \multicolumn{3}{c}{Median}   \\
\cmidrule{2-4}\cmidrule{6-8}
\multirow{-2}{*}{Model} & Total & $I_\text{spa}$ & $\eta_a$ & & Total & $I_\text{spa}$ & $\eta_a$ \\
\cmidrule{1-4}\cmidrule{6-8}
SEM & 18.0  & 16.4    & 19.6       &  & 8.8    & 7.2      & 10.7     \\
FNN  & 14.7   & 8.4     & 20.0   &  & 5.8   & 5.1      & 8.0   \\               
\bottomrule
\end{tabular}
\end{table}

Table~\ref{tab:APE_FNN_SEM_comparison} represents the mean and median values of the total absolute percentage error distribution of the two models (Fig.~\ref{fig:error_distribution}, left) and compares accuracy of the anode efficiency $\eta_a$ and anode specific impulse $I_{spa}$ predictions. It indicates that for the FNN ensemble, the mean APE is significantly lower for the specific impulse (8.4$\%$ versus 16.4$\%$) but slightly higher for the efficiency (20.0$\%$ versus 19.6$\%$). However, the median APE for the FNN ensemble is lower for both specific impulse and efficiency (5.1$\%$ versus 7.2$\%$ and 8.0$\%$ versus 10.7$\%$ respectively). The reason for the difference in the trends for the mean and median APE values are those thrusters with outlying parameter values in the database. The median APE is mainly stable in the presence of outliers, whereas the mean APE is highly influenced by them. Thus, as can be seen from Table~\ref{tab:APE_FNN_SEM_comparison}, in the parametric domain covered by the database, the prediction accuracy of the FNN ensemble is higher than that of the SEM \cite{Shagayda2015}.

\begin{table}[hbt!]
\caption{\label{tab:APE_folds} APE for different fold combinations, $\%$}
\centering
\begin{tabular}{cccccccclcccccc}
\toprule
\multirow{3}{*}{\begin{tabular}[c]{@{}c@{}}Test \\ index\end{tabular}} & \multirow{3}{*}{\begin{tabular}[c]{@{}c@{}}Training  \\ indices\end{tabular}} & \multicolumn{6}{c}{Mean} &  & \multicolumn{6}{c}{Median} \\
\cmidrule{3-8}\cmidrule{10-15}
 &  & \multicolumn{2}{c}{Total} & \multicolumn{2}{c}{$I_\text{spa}$} & \multicolumn{2}{c}{$\eta_a$} &  & \multicolumn{2}{c}{Total} & \multicolumn{2}{c}{$I_\text{spa}$} & \multicolumn{2}{c}{$\eta_a$} \\
\cmidrule{3-8}\cmidrule{10-15}
 &  & Train & Test & Train & Test & Train & Test &  & Train & Test & Train & Test & Train & Test \\
\midrule
1 & 2--5       & 8.5  & 23.1 & 5.3 & 15.6 & 11.6 & 30.5 &  & 3.9 & 6.4 & 2.9 & 5.9 & 5.0 & 7.5 \\
2 & 1, 3--5    & 10.2 & 15.8 & 6.2 & 10.6 & 14.2 & 21.0 &  & 3.2 & 8.3 & 2.4 & 7.3 & 4.1 & 12.1 \\
3 & 1, 2, 4, 5 & 11.4 & 13.8 & 7.1 & 6.8  & 15.7 & 20.8 &  & 4.9 & 6.2 & 3.7 & 4.5 & 6.5 & 9.3 \\
4 & 1--3, 5    & 10.6 & 11.7 & 6.8 & 7.8  & 14.4 & 15.7 &  & 4.3 & 5.1 & 3.4 & 5.1 & 5.8 & 6.2 \\
5 & 1--4       & 9.4  & 9.2  & 5.7 & 6.1  & 13.1 & 12.2 &  & 3.5 & 3.3 & 2.7 & 2.6 & 4.7 & 4.9 \\
\bottomrule
\end{tabular}
\end{table}

The machine learning approach not only predicts the parameters of HETs but can also be used to analyze the data. The distribution of the errors for the FNN ensemble for different training and test samples (Table~\ref{tab:APE_folds}) suggests that there is a strong variance in the data. For example, when training the model on samples 2--5 and testing on sample 1, this model has large  errors for the test sample but low errors for the training sample. Moreover, in this particular case, the total mean APE for the test sample is higher than for all the other test samples, and the total mean APE for the training sample is lower than for all the other training samples. This suggests that there are both thrusters with outliers and those without.

\begin{figure}[ht!]
\centering
\includegraphics[width=1.\textwidth]{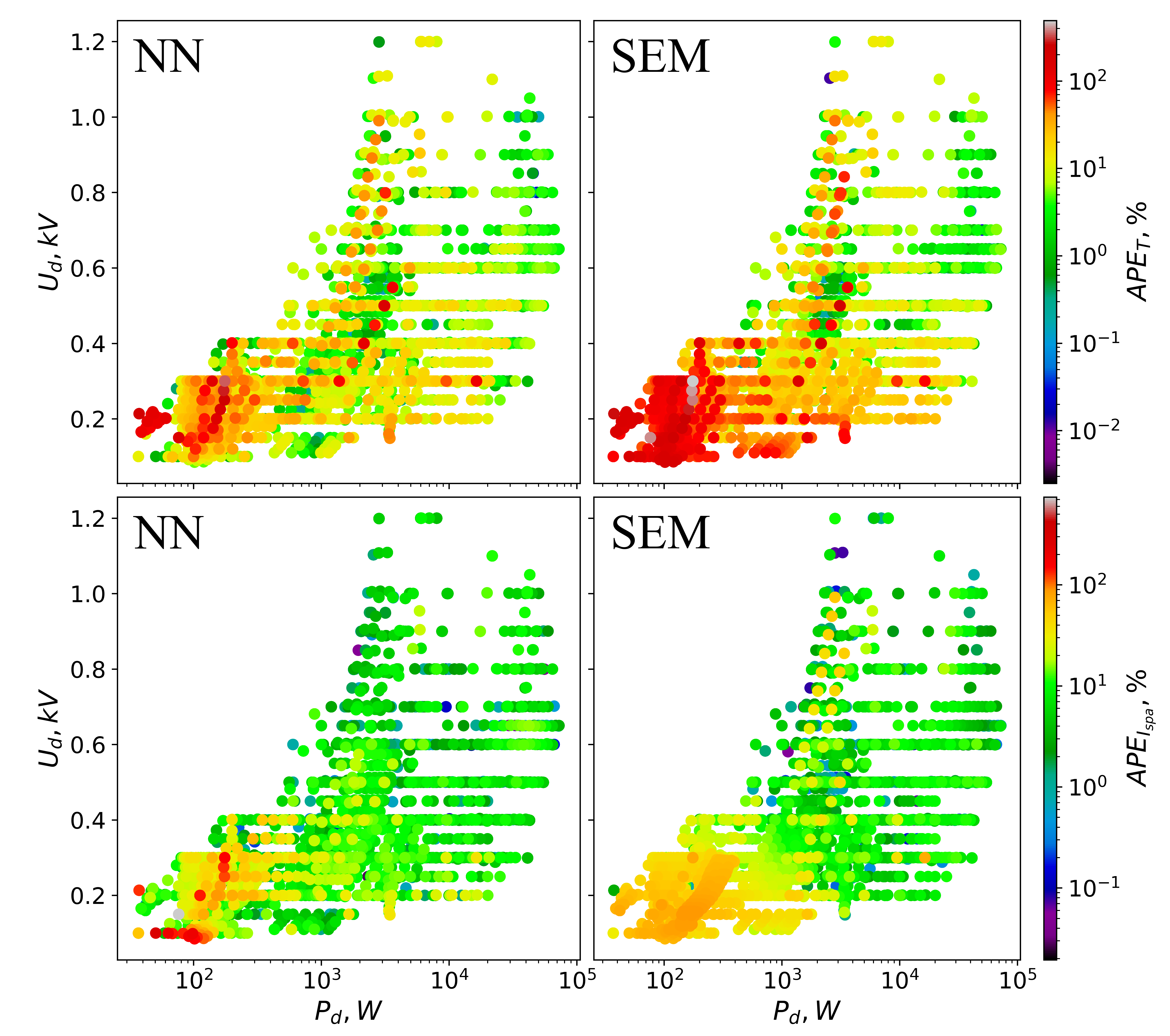}
\caption{Location of the thrust and anode specific impulse maximum errors for the FNN ensemble and SEM \cite{Shagayda2015} in the discharge parameter space.}
\label{fig:APE_localization_Pd_Ud}
\end{figure}

Figure ~\ref{fig:APE_localization_Pd_Ud} shows the distribution of the thrust $T$ and anode specific impulse $I_{spa}$ prediction error of the FNN ensemble and SEM \cite{Shagayda2015} for different experimental values of the discharge power and voltage in the database. The largest errors are mainly concentrated in the region with a low discharge power ($P_d < 400$~W) for both models. The area with maximum thrust and anode specific impulse errors for the SEM \cite{Shagayda2015} is higher than one for the FNN ensemble. Moreover, the maximum thrust error for the SEM \cite{Shagayda2015} is higher than one for the FNN ensemble. However, the maximum specific impulse error for the FNN ensemble is higher than for the SEM \cite{Shagayda2015}. The errors mainly depend on two factors. The first is the difference between the model's prediction and sample entries, and the second is the spread of the experimental data around some average value. In other words, in the case of the FNN ensemble, a high error shows that thrusters created by different developers with similar dimensions in similar discharge modes have significantly different output characteristics. Three factors affect this scatter in the parameters of the low-power HET operation.

First, in HET development, a common technique is to scale an existing and successful thruster up or down to obtain a new configuration with the required operating parameters. However, for small HETs, there are some additional difficulties that require special approaches. It can be assumed that since the engineering methods for the development of small HETs are less proven than for medium-sized and large ones, the devices constructed may have very different operating efficiencies, which may explain the wide dispersion of their parameters.

Second, in a firing test of a low-power HET, the low values of the thrust and propellant flow rate must be measured accurately. However, the experimental conditions have a huge influence on the operation of low-power HETs, which potentially degrades measurement precision.

Third, novice research teams often start experimenting in the low-power range. This makes us somewhat skeptical about these results, and thus we expect significant variance in these data.
The combination of these three factors could explain the observed high variance in the performance of low-power thrusters.

\subsection{Dependency of thruster performance on the operating parameters}

\begin{figure}[ht!]
\centering
\includegraphics[width=1.0\textwidth]{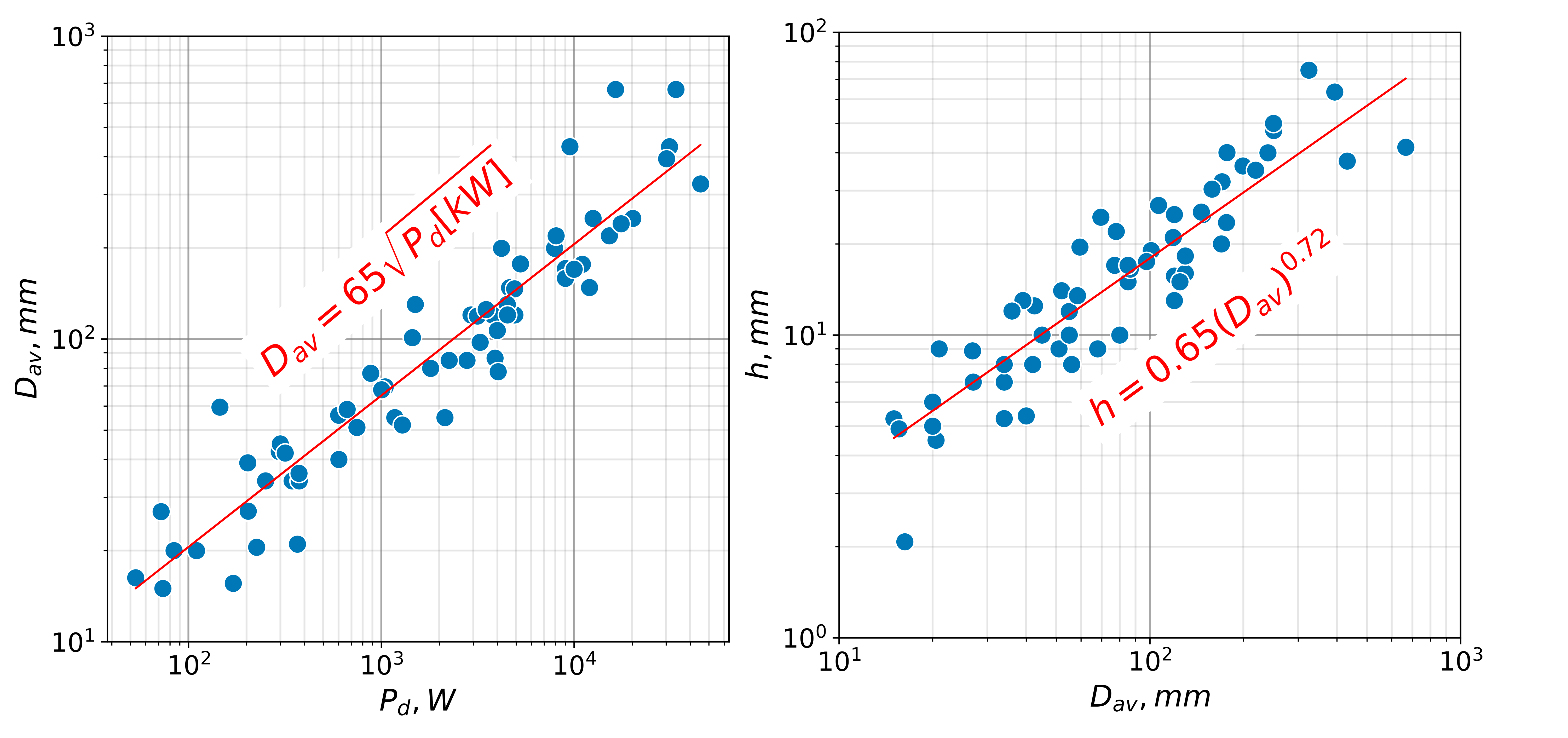}
\caption{Scattering of the nominal power $P_d$ and the mean diameter $D_\text{av}$ and height $h$ of the discharge channel, and approximations of $D_\text{av}$ and $h$.}
\label{fig:D_av and h approximations}
\end{figure}

The predictions for the FNN ensemble were compared with those of the reference model \cite{Shagayda2015}.
The SEM \cite{Shagayda2015} assumes the unambiguous dependence of the mean diameter and height of the discharge channel on the discharge power. The straight lines in Fig.~\ref{fig:D_av and h approximations} show the form of these dependencies. However, the same Fig.~\ref{fig:D_av and h approximations} also demonstrates a scatter of these parameters. In the FNN ensemble, we set the channel height and diameter as independent input features, which allowed us to study how the discharge channel geometry affects the HET efficiency. Note that it was impossible when using the SEM model \cite{Shagayda2015}. However, to unify the assumptions of the two models for a fair comparison, the original assumption about the solid dependency between the discharge channel geometry and discharge power was enforced in this section for the FNN ensemble as well. The corresponding dependencies were regressed from the actual distributions, as illustrated in Fig.~\ref{fig:D_av and h approximations}.

\begin{figure}[ht!]
\centering
\includegraphics[width=1.0\textwidth]{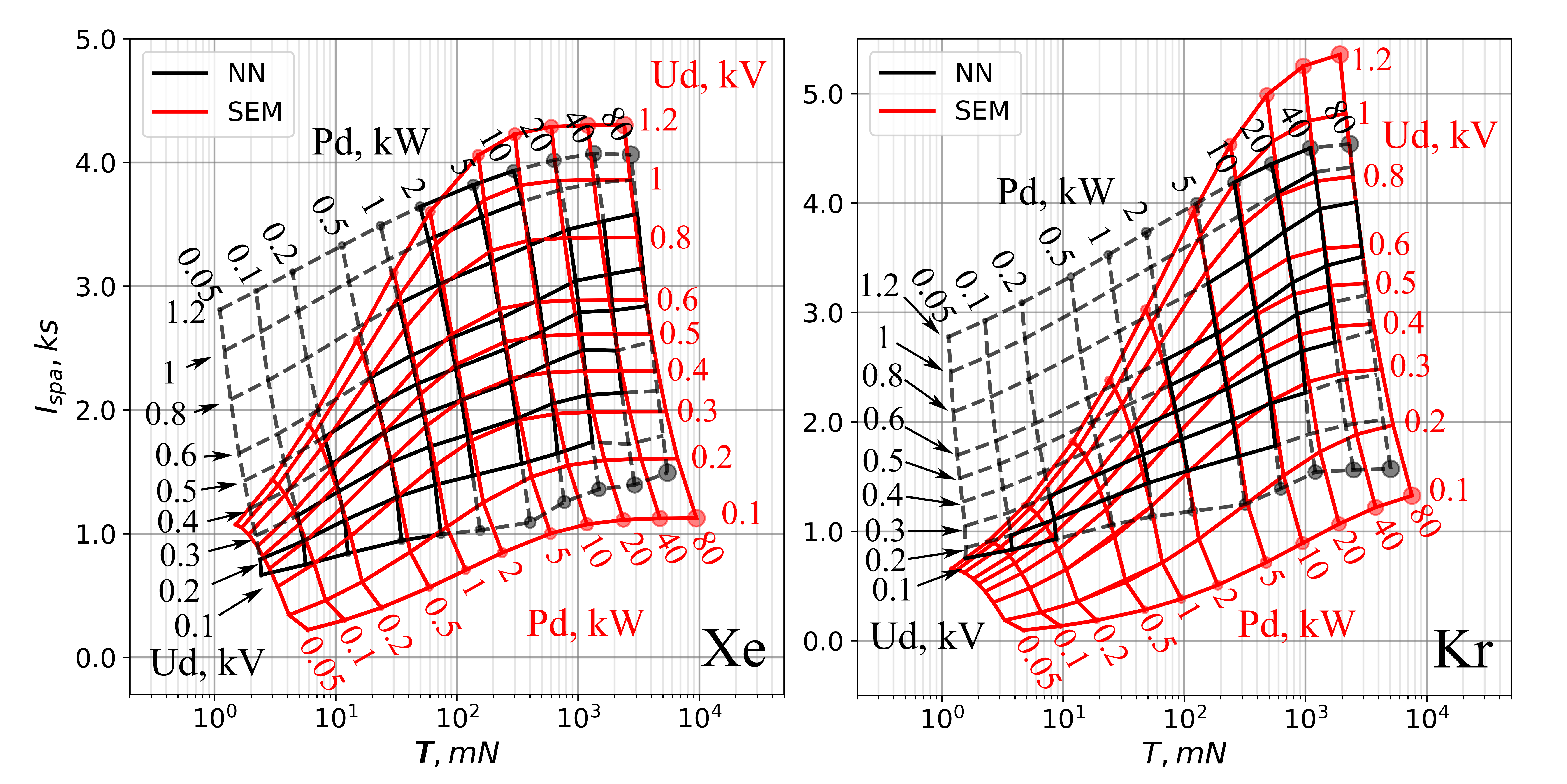}
\caption{Comparison of the FNN ensemble (NN) and SEM \cite{Shagayda2015} predictions for the explicit dependence of $D_\text{av}$ on $P_d$ and $h$ on $D_\text{av}$ (Fig.~\ref{fig:D_av and h approximations}).}
\label{fig:FNNvsSEM}
\end{figure}

Fig.~\ref{fig:D_av and h approximations} shows the distributions of data points in the mean diameter $D_\text{av}$ and discharge power $P_d$ space (left) and in the mean diameter $D_\text{av}$ and height $h$ space (right). These distributions can be approximated by the power laws shown in the figure. When comparing the predictions of the FNN ensemble and SEM \cite{Shagayda2015}, the values of the independent input parameters were set with values from these power laws.
This comparison in this section focuses on xenon and krypton propellants. The thruster type was also fixed as HET because the 2012 version of the database used for tuning the model \cite{Shagayda2015} contains data only for HET thrusters.

\begin{figure}[ht!]
\centering
\includegraphics[width=1.0\textwidth]{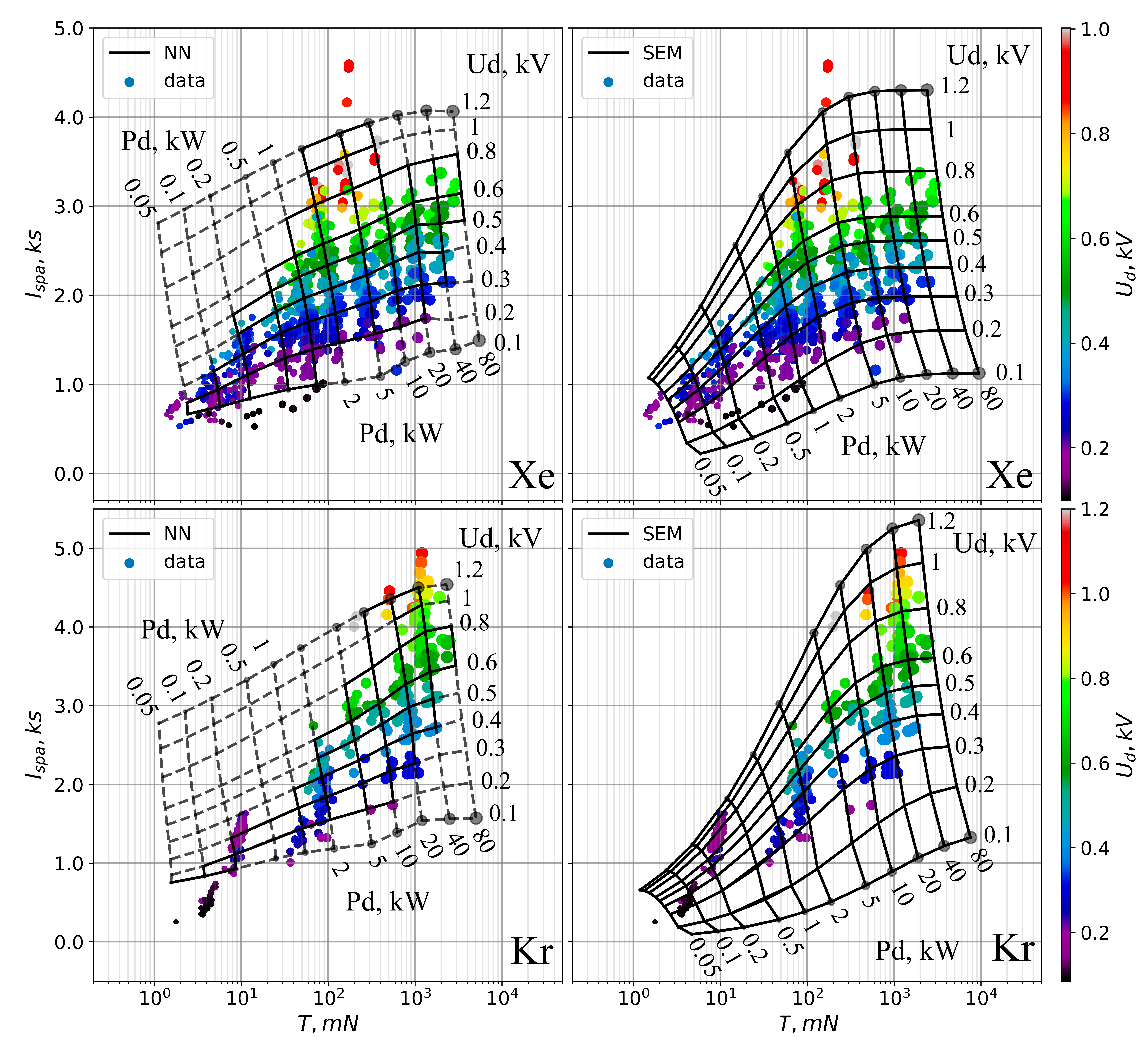}
\caption{Predictions of the FNN ensemble (NN) and SEM \cite{Shagayda2015} for the explicit dependence of $D_\text{av}$ on $P_d$ and $h$ on $D_\text{av}$ (Fig.~\ref{fig:D_av and h approximations}), and data scatter.}
\label{fig:main_result}
\end{figure}

Fig.~\ref{fig:FNNvsSEM} presents the predicted dependencies between the input parameters $U_d$ and $P_d$ and the predicted output performance $I_\text{spa}$ and $T$ [via the predicted $\eta_a$; Eq.~(\ref{Isp and eta})] for Xe and Kr propellants. Note that here the discharge channel geometry is constrained by the $P_d$ approximation, as discussed above. For each model, the pair of predicted values ($I_\text{spa}$, $T$) is presented for pairs of input parameters ($U_d$, $P_d$) taken from the regular 2D grid.
There are reference points (gray) on the lines for the minimum and maximum voltages in the grids for the FNN ensemble and for SEM \cite{Shagayda2015}.

In Fig.~\ref{fig:FNNvsSEM}, the red curves correspond to the predictions of the SEM \cite{Shagayda2015}. The dependencies predicted by the FNN ensemble model are presented by black lines. Note, that  reliable predictions by the FNN model are limited to the domain of the models in the database used for training. These predictions are shown as solid black lines. Predictions for the model obtained for parameters beyond the training domain are shown as dashed lines, and these should be treated cautiously.

Fig.~\ref{fig:main_result} shows the distribution of the database entries and grids for the FNN ensemble (top left and bottom left) and the SEM (top right and bottom right) predictions for Xe and Kr propellants. The color of the database points corresponds to the discharge voltage, as indicated by the color bars. The sizes of the points reflect the discharge power.

Fig.~\ref{fig:FNNvsSEM} indicates that the best agreement is in the ranges 2--5~kW and 300--800~V for xenon and 5--10~kW and 300--400~V for krypton. The experimental data in this range are mainly for well-developed thrusters with consistently good operating efficiency. 

According to Fig~\ref{fig:main_result}, the FNN ensemble’s predictions are much more accurate in the 0.2--2~kW, 200~V, and 0.2--1~kW, 300~V regions for xenon. The specific impulse predicted by the SEM \cite{Shagayda2015} in these regions turns out to be lower, while the thrust is higher, which is weakly consistent with the data. Also, predictions of the FNN ensemble appear to be slightly more accurate than those of the SEM in the vicinity of the following points: 2~kW and 300~V, 20~kW and 300~V, and 5~kW and 600~V for krypton. However, the SEM performs better between 1000--1200~V, 20~kW for krypton. Table~\ref{tab:APE_FNN_SEM_comparison} gives a numerical comparison of the accuracy of the two models.

At the extremes of the operating range of the HETs, especially in the low-power area, there are significant discrepancies in the predictions of the models. There are two primary reasons for these discrepancies: imperfections in the SEM \cite{Shagayda2015} and a lack of data near the edges of the HET operating range. A degradation of the quality of the FNN ensemble model is expected in parameter regions beyond the model training domain. If predictions need to be extrapolated, the SEM \cite{Shagayda2015} may give better predictions than the machine learning model.

There are practical uses of the dependencies of the thrust and specific impulse on the voltage and discharge power, as shown in  Fig.~\ref{fig:FNNvsSEM} and Fig.~\ref{fig:main_result}. They can be applied to make a preliminary estimate of the HET operating parameters (discharge power and voltage) that achieve the required thrust and specific impulse, which can be useful in the development and design of HETs.

\subsection{Dependency of the output parameters on the operating mode and discharge channel geometry}

\begin{figure}[htb!]
\centering
\includegraphics[width=1.\textwidth]{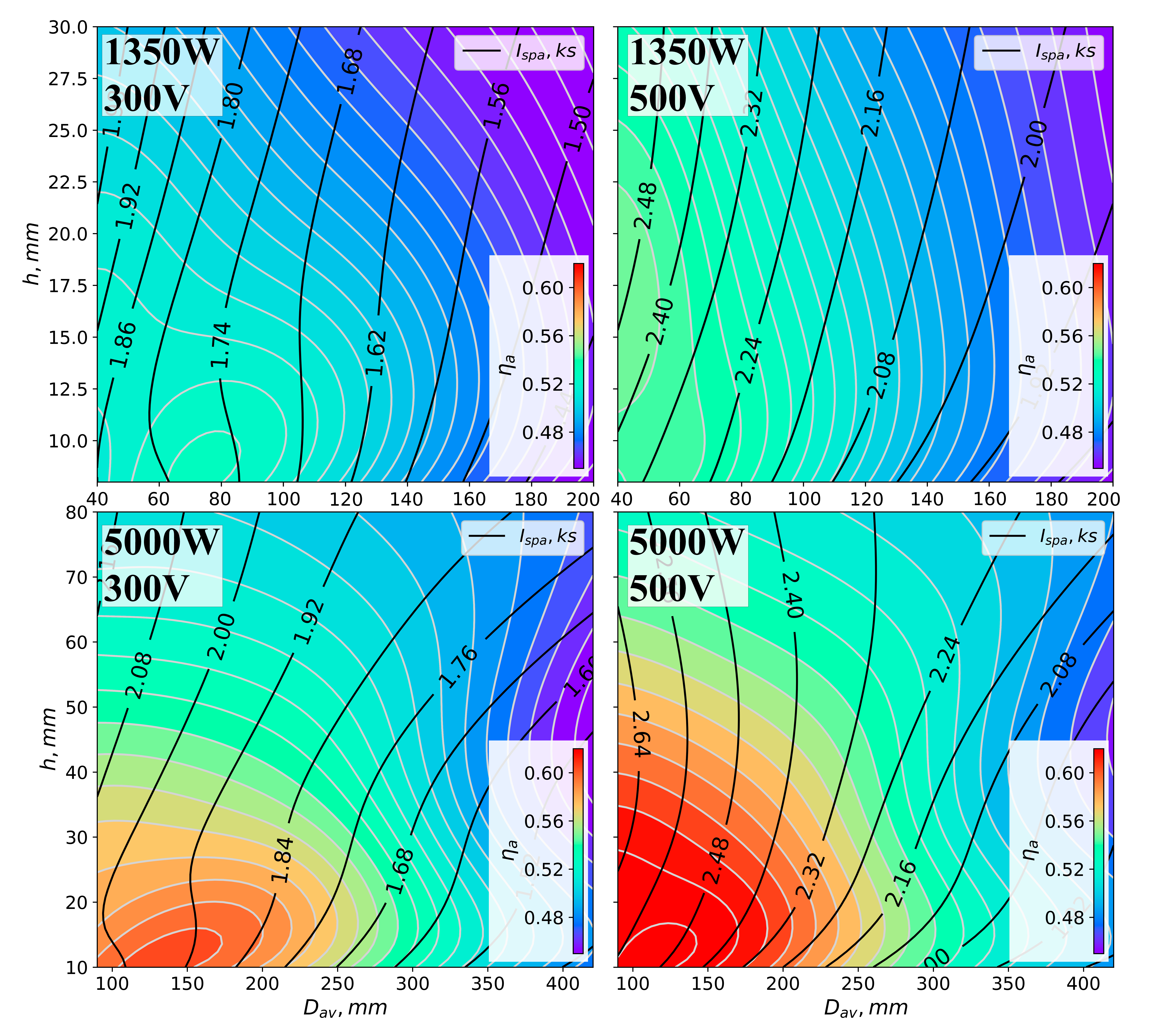}
\caption{Predictions from the FNN ensemble of the anode efficiency $\eta_a$ and anode specific impulse $I_\text{spa}$ as functions of the discharge channel geometric parameters for the indicated discharge voltages and powers.}
\label{fig:geometry_variations}
\end{figure}

As noted above, the new model based on the FNN ensemble treats the discharge channel height and diameter as input parameters. Figure~\ref{fig:geometry_variations} shows the anode efficiency (colored contours) and anode specific impulse (black contour lines) predicted by the FNN ensemble model as functions of the discharge channel height and mean diameter for several values of the voltage and discharge power for HET thrusters with xenon propellant. 

The dependencies have similar trends for different operating modes. This similarity may be due to the specific physical mechanisms affecting the statistics of the data, which allows the FNN ensemble to learn them implicitly. However, at the current stage of this work, these mechanisms cannot be confidently identified. The local optimum of the anode efficiency may have the following qualitative explanation. If the thrusters are too large for the selected discharge mode, they may suffer from low ion-forming efficiency. In contrast, thrusters that are too small have increased wall losses and may experience overheating. The first reason may also explain the decrease in the anode specific impulse with an increase of the discharge channel diameter. Ultimately, a rigorous physically based inference requires reliable experimental and numerical data.

\section{Conclusion}\label{sec:v}

For the first time, machine learning methods have been used to create a HET scaling model. The new model, which is based on an FNN ensemble, was built from an extensive database of HET parameters collected from publications, conference proceedings, and Ph.D. theses. It predicts the performance of a thruster using discharge parameters, propellant type, and discharge channel construction. This model complements the SEM \cite{Shagayda2015}. 

The model has two advantages: 1) It is more accurate in the domain of the input parameters corresponding to actual database entries; 2) it can be built without needing to make specific assumptions about $\eta_I$ and $\eta_U$ and assuming no explicit dependence of $D_\text{av}$ on $P_d$ and $h$ on $D_\text{av}$.
In the parameter domain, covered by the database,  the model's predictions are more accurate than predictions of the SEM \cite{Shagayda2015}. A natural limitation of the new model, inherited from the machine learning approach per se, is its poor extrapolation ability. Thus, the predictions of this model can be suitable when creating new thrusters, the parameters of which lie within the scatter of the database. In turn, if parameters of new thrusters lie outside this domain, extrapolation of predictions is necessary. In this case, the SEM \cite{Shagayda2015} may, indeed, give better predictions.

We believe that further development of HET scaling models will be in the synthesis of physically based theories and methods of machine learning. For example, in the previously developed SEM \cite{Shagayda2015}, the current and voltage utilization coefficients ($\eta_I$ and $\eta_U$) were fixed because their dependencies on the operating modes of HETs were unknown. Using machine learning techniques to search for these dependencies looks attractive. Modern neural nets can handle much more complicated data structures than the ones presented, so it is feasible to include additional features as the input of a machine learning model, if the required data are available. 

\bibliography{main}

\begin{thebibliography}{26}
\newcommand{\enquote}[1]{``#1''}
\providecommand{\natexlab}[1]{#1}
\providecommand{\url}[1]{\texttt{#1}}
\providecommand{\urlprefix}{URL }
\expandafter\ifx\csname urlstyle\endcsname\relax
  \providecommand{\doi}[1]{\discretionary{}{}{}https://doi.org/#1}\else
  \providecommand{\doi}[1]{\discretionary{}{}{}\urlstyle{rm}\url{https://doi.org/#1}}\fi

\bibitem[{Morozov and Savelyev(2000)}]{Morozov2000}
Morozov, A.~I., and Savelyev, V.~V., \enquote{Fundamentals of Stationary Plasma
  Thruster Theory,} \emph{Reviews of Plasma Physics}, Springer {US}, 2000, pp.
  203--391.
\newblock \doi{10.1007/978-1-4615-4309-1_2}.

\bibitem[{Kim(1998)}]{Kim1998-ro}
Kim, V., \enquote{Main Physical Features and Processes Determining the
  Performance of Stationary Plasma Thrusters,} \emph{Journal of Propulsion and
  Power}, Vol.~14, No.~5, 1998, pp. 736--743.
\newblock \doi{10.2514/2.5335}.

\bibitem[{Lev et~al.(2019)Lev, Myers, Lemmer, Kolbeck, Koizumi, and
  Polzin}]{Lev2019-li}
Lev, D., Myers, R.~M., Lemmer, K.~M., Kolbeck, J., Koizumi, H., and Polzin, K.,
  \enquote{The Technological and Commercial Expansion of Electric Propulsion,}
  \emph{Acta Astronautica}, Vol. 159, 2019, pp. 213--227.
\newblock \doi{10.1016/j.actaastro.2019.03.058}.

\bibitem[{Boeuf(2017)}]{Boeuf2017-vr}
Boeuf, J.-P., \enquote{Tutorial: Physics and Modeling of Hall Thrusters,}
  \emph{Journal of Applied Physics}, Vol. 121, No.~1, 2017, p. 011101.
\newblock \doi{10.1063/1.4972269}.

\bibitem[{Kaganovich et~al.(2020)Kaganovich, Smolyakov, Raitses, Ahedo,
  Mikellides, Jorns, Taccogna, Gueroult, Tsikata, Bourdon, Boeuf, Keidar,
  Powis, Merino, Cappelli, Hara, Carlsson, Fisch, Chabert, Schweigert, Lafleur,
  Matyash, Khrabrov, Boswell, and Fruchtman}]{Kaganovich2020}
Kaganovich, I.~D., Smolyakov, A., Raitses, Y., Ahedo, E., Mikellides, I.~G.,
  Jorns, B., Taccogna, F., Gueroult, R., Tsikata, S., Bourdon, A., Boeuf,
  J.-P., Keidar, M., Powis, A.~T., Merino, M., Cappelli, M., Hara, K.,
  Carlsson, J.~A., Fisch, N.~J., Chabert, P., Schweigert, I., Lafleur, T.,
  Matyash, K., Khrabrov, A.~V., Boswell, R.~W., and Fruchtman, A.,
  \enquote{Physics of E{\hspace{0.167em}}{\texttimes}{\hspace{0.167em}}B
  Discharges Relevant to Plasma Propulsion and Similar Technologies,}
  \emph{Physics of Plasmas}, Vol.~27, No.~12, 2020, p. 120601.
\newblock \doi{10.1063/5.0010135}.

\bibitem[{Morozov and Melikov(1974)}]{Morozov1974-vt}
Morozov, A.~I., and Melikov, I.~V., \enquote{On Similarity of Processes in
  Plasma Accelerators with Closed Electron Drift in Conditions of Ionization,}
  \emph{Zhurnal Tekhnicheskoj Fiziki}, Vol.~44, No.~3, 1974, pp. 544--548.

\bibitem[{Bugrova et~al.(1991)Bugrova, Maslennikov, and
  Morozov}]{A_I_Bugrova_N_A_Maslennikov_A_I_Morozov1991-vr}
Bugrova, A.~I., Maslennikov, N.~A., and Morozov, A.~I., \enquote{Similarity
  Laws of the {ACDE} Integral Characteristics,} \emph{Journal of Technical
  Physics}, Vol.~61, No.~6, 1991, pp. 45--51.

\bibitem[{Fruchtman et~al.(1997)Fruchtman, Fisch, Ashkenazy, and
  Raitses}]{fruchtman_scaling_laws}
Fruchtman, A., Fisch, N.~J., Ashkenazy, J., and Raitses, Y., \enquote{Scaling
  Laws for Hall Thruster Performance,} \emph{Proceedings of the 25th
  International Electric Propulsion Conference}, Vol.~8, Cleveland, OH, 1997,
  pp. 24--8.

\bibitem[{Ahedo and Gallardo(2003)}]{E_Ahedo_undated-xh}
Ahedo, E., and Gallardo, J., \enquote{Scaling Down Hall Thrusters,}
  \emph{Proceedings of the 28th International Electric Propulsion Conference},
  Toulouse, France, 2003, p. 104.

\bibitem[{Andrenucci et~al.(2003)Andrenucci, Biagioni, Marcuccio, Paganucci,
  and
  Tobak}]{M_Andrenucci_L_Biagioni_S_Marcuccio_F_Paganucci_and_M_Tobak_undated-xl}
Andrenucci, M., Biagioni, L., Marcuccio, S., Paganucci, F., and Tobak, M.,
  \enquote{Fundamental Scaling Laws for Electric Propulsion Concepts. Part 1:
  Hall Effect Thrusters,} \emph{Proceedings of the 28th International Electric
  Propulsion Conference}, Toulouse, France, 2003.

\bibitem[{Ashkenazy et~al.(2005)Ashkenazy, Shitrit, and
  Appelbaum}]{J_Ashkenazy_S_Shitrit_and_G_Appelbaum_undated-hh}
Ashkenazy, J., Shitrit, S., and Appelbaum, G., \enquote{Hall Thruster
  Modifications for Reduced Power Operation,} \emph{Proceedings of the 29th
  International Electric Propulsion Conference}, Princeton, NJ, 2005.
\newblock \doi{10.13140/2.1.2075.3280}.

\bibitem[{Andrenucci et~al.(2005)Andrenucci, Battista, and
  Piliero}]{M_Andrenucci_F_Battista_and_P_Piliero_undated-oh}
Andrenucci, M., Battista, F., and Piliero, P., \enquote{Hall Thruster Scaling
  Methodology,} \emph{Proceedings of the 29th International Electric Propulsion
  Conference}, Princeton, NJ, 2005.

\bibitem[{Daren et~al.(2005)Daren, Yongjie, and Zhi}]{Daren2005-mi}
Daren, Y., Yongjie, D., and Zhi, Z., \enquote{Improvement on the Scaling Theory
  of the Stationary Plasma Thruster,} \emph{Journal of Propulsion and Power},
  Vol.~21, No.~1, 2005, pp. 139--143.
\newblock \doi{10.2514/1.5901}.

\bibitem[{Battista et~al.(2007)Battista, De~Marco, and
  Misuri}]{F_Battista_E_A_D_Marco_and_T_Misuri_undated-xm}
Battista, F., De~Marco, E.~A., and Misuri, T., \enquote{A Review of the Hall
  Thruster Scaling Methodology,} \emph{Proceedings of the 30th International
  Electric Propulsion Conference}, Florence, Italy, 2007.

\bibitem[{Misuri and Andrenucci(2008)}]{Misuri2008-cd}
Misuri, T., and Andrenucci, M., \enquote{{HET} Scaling Methodology: Improvement
  and Assessment,} \emph{44th {AIAA}/{ASME}/{SAE}/{ASEE} Joint Propulsion
  Conference and Exhibit}, American Institute of Aeronautics and Astronautics,
  2008.
\newblock \doi{10.2514/6.2008-4806}.

\bibitem[{Shagayda and Gorshkov(2013)}]{AA_Shagayda_OA_Gorshkov2013-gc}
Shagayda, A., and Gorshkov, O., \enquote{{Hall-Thruster} Scaling Laws,}
  \emph{Journal of Propulsion and Power}, Vol.~29, No.~2, 2013, pp. 466--474.
\newblock \doi{10.2514/1.B34650}.

\bibitem[{Shagayda(2015)}]{Shagayda2015}
Shagayda, A.~A., \enquote{On Scaling of Hall Effect Thrusters,} \emph{IEEE
  Transactions on Plasma Science}, Vol.~43, No.~1, 2015, pp. 12--28.
\newblock \doi{10.1109/TPS.2014.2315851}.

\bibitem[{Lee et~al.(2019)Lee, Kim, Lee, Kim, Doh, Lee, and Choe}]{Choe2019}
Lee, E., Kim, Y., Lee, H., Kim, H., Doh, G., Lee, D., and Choe, W.,
  \enquote{Scaling Approach for Sub-Kilowatt Hall-Effect Thrusters,}
  \emph{Journal of Propulsion and Power}, Vol.~35, No.~6, 2019, pp. 1073--1079.
\newblock \doi{10.2514/1.B37424}.

\bibitem[{Olano~Garcia et~al.(2020)Olano~Garcia, Tang, and
  Ren}]{Olano_Garcia2020-ds}
Olano~Garcia, A., Tang, H., and Ren, J., \enquote{Scaling Model for {SPT} and
  {TAL} Thrusters,} \emph{IEEE Transactions on Plasma Science}, Vol.~48, No.~1,
  2020, pp. 86--98.
\newblock \doi{10.1109/TPS.2019.2958187}.

\bibitem[{Fuchigami et~al.(2017)Fuchigami, Egawa, Morita, and
  Yamamoto}]{IEPC-2017-453}
Fuchigami, H., Egawa, Y., Morita, T., and Yamamoto, N., \enquote{Prediction of
  the Thruster Performance in Hall Thrusters Using Neural Network,}
  \emph{Proceedings of the 35th International Electric Propulsion Conference},
  Atlanta, GA, 2017.

\bibitem[{Mörtl et~al.(2019)Mörtl, Knoll, Williams, Shaw, Argyriou, Zamattio,
  and Pugliese}]{IEPC-2019-501}
Mörtl, M., Knoll, A., Williams, V., Shaw, P., Argyriou, V., Zamattio, J., and
  Pugliese, L., \enquote{Enhancing Hall Effect Thruster Simulations with Deep
  Recurrent Networks,} \emph{Proceedings of the 36th International Electric
  Propulsion Conference}, Vienna, Austria, 2019.

\bibitem[{Williams et~al.(2019)Williams, Argyriou, Montag, Herdrich, Knoll, and
  Mörtl}]{IEPC-2019-169}
Williams, V., Argyriou, V., Montag, C., Herdrich, G., Knoll, A., and Mörtl,
  M., \enquote{Development of PPTNet a Neural Network for the Rapid Prototyping
  of Pulsed Plasma Thrusters,} \emph{Proceedings of the 36th International
  Electric Propulsion Conference}, Vienna, Austria, 2019.

\bibitem[{Paszke et~al.(2019)Paszke, Gross, Massa, Lerer, Bradbury, Chanan,
  Killeen, Lin, Gimelshein, Antiga, Desmaison, Kopf, Yang, DeVito, Raison,
  Tejani, Chilamkurthy, Steiner, Fang, Bai, and Chintala}]{Paszke2019}
Paszke, A., Gross, S., Massa, F., Lerer, A., Bradbury, J., Chanan, G., Killeen,
  T., Lin, Z., Gimelshein, N., Antiga, L., Desmaison, A., Kopf, A., Yang, E.,
  DeVito, Z., Raison, M., Tejani, A., Chilamkurthy, S., Steiner, B., Fang, L.,
  Bai, J., and Chintala, S., \enquote{PyTorch: An Imperative Style,
  High-Performance Deep Learning Library,} \emph{Advances in Neural Information
  Processing Systems 32}, Curran Associates, Inc., 2019, pp. 8024--8035.

\bibitem[{Ioffe and Szegedy(2015)}]{ioffe2015}
Ioffe, S., and Szegedy, C., \enquote{Batch Normalization: Accelerating Deep
  Network Training by Reducing Internal Covariate Shift,} \emph{Proceedings of
  the 32nd International Conference on Machine Learning}, Vol.~37, Lille,
  France, 2015, p. 448–456.

\bibitem[{Pedregosa et~al.(2011)Pedregosa, Varoquaux, Gramfort, Michel,
  Thirion, Grisel, Blondel, Prettenhofer, Weiss, Dubourg, Vanderplas, Passos,
  Cournapeau, Brucher, Perrot, and Duchesnay}]{Pedregosa2011-nh}
Pedregosa, F., Varoquaux, G., Gramfort, A., Michel, V., Thirion, B., Grisel,
  O., Blondel, M., Prettenhofer, P., Weiss, R., Dubourg, V., Vanderplas, J.,
  Passos, A., Cournapeau, D., Brucher, M., Perrot, M., and Duchesnay, E.,
  \enquote{Scikit-learn: Machine Learning in {P}ython,} \emph{Journal of
  Machine Learning Research}, Vol.~12, 2011, pp. 2825--2830.

\bibitem[{Lema\^{\i}tre et~al.(2017)Lema\^{\i}tre, Nogueira, and
  Aridas}]{10.5555/3122009.3122026}
Lema\^{\i}tre, G., Nogueira, F., and Aridas, C.~K., \enquote{Imbalanced-Learn:
  A Python Toolbox to Tackle the Curse of Imbalanced Datasets in Machine
  Learning,} \emph{Journal of Machine Learning Research}, Vol.~18, No.~1, 2017,
  p. 559–563.

\end{thebibliography}

\end{document}